\documentstyle[epsfig,prl,aps]{revtex}
\title{Concentration Dependent Sedimentation of Colloidal Rods}
\author{Z. Dogic$^{\ast}$, A.P. Philipse$^{\dagger}$, S. Fraden$^{\ast}$,
J.K.G. Dhont$^{\dagger}$ $^{\ddagger}$ $^{\star}$}
\address{$^{\ast}$Complex Fluid Group, Department of Physics, Brandeis University,
Waltham MA 02454, \linebreak $^{\dagger}$van't Hoff Laboratory,
Utrecht University, Padualaan8, 3584 CH Utrecht, The Netherlands,
$^{\ddagger}$Forschungzentrum J\"{u}lich, IFF, Weiche Materie,
52425 J\"{u}lich, Germany}

\date{\today}

\renewcommand{\r}{{\mathbf r}}
\newcommand{\T}{{\mathbf T}}
\newcommand{\f}{{\mathbf f}}
\newcommand{\F}{{\mathbf F}}
\newcommand{\vv}{{\mathbf v}}
\newcommand{\uu}{{\mathbf u}}
\begin{document}

\maketitle
\begin{abstract}

In the first part of this paper, an approximate theory is developed
for the leading order concentration dependence of the sedimentation
coefficient for rod-like colloids/polymers/macromolecules. To first
order in volume fraction $\varphi$ of rods, the sedimentation
coefficient is written as $1+\alpha \varphi$. For large aspect
ratio L/D (L is the rod length, D it's thickness) $\alpha$ is found
to very like $\propto \left( \frac{L}{D} \right)^2/\log \left(
\frac{L}{D} \right)$. This theoretical prediction is compared to
experimental results. In the second part, experiments on {\it
fd}-virus are described, both in the isotropic and nematic phase.
First order in concentration results for this very long and thin
(semi-flexible) rod are in agreement with the above theoretical
prediction. Sedimentation profiles for the nematic phase show two
sedimentation fronts. This result indicates that the nematic phase
becomes unstable with the respect to isotropic phase during
sedimentation.
\end{abstract}

\section{Introduction}

There is extensive literature concerned with sedimentation
behaviour of spherically shaped colloidal particles (for a review
see ref. [1]). Essentially exact predictions can be made for the
sedimentation velocity of spherical colloids to first order in
concentration [2]. For non-spherical colloids a similar exact
prediction is non-existent. The only attempt to calculate the first
order concentration dependence of the sedimentation velocity for
rod like colloids we are aware of is due to Peterson [3]. This
theory is based on approximate, orientationally pre-averaged
hydrodynamic interactions between the colloidal rods and a rather
crude estimate of certain multiple integrals that represent the
ensemble averaged velocity. As yet there are no accurate
expressions for hydrodynamic interaction tensors for rods. In the
first part of the present paper, in section II, we calculate these
interaction tensors in a mean-field approximation. In section III
we use this approximate expression for the hydrodynamic interaction
functions to derive an explicit expression for the first order in
concentration coefficient of the sedimentation velocity as a
function of the aspect ratio of the rods. This expression is found
to agree remarkably well with Peterson's result for aspect ratios
less than about $30$. For larger aspect ratios our result for the
first order in concentration coefficient is much larger then
Peterson's prediction. In the second part of this paper, section
IV, sedimentation experiments on {\it fd}-virus are discussed.
Experiments are done at low concentration to find the first order
concentration dependence, which is compared to the theory mentioned
above. In addition, sedimentation experiments at larger
concentrations, including the nematic phase are performed.

\section{Hydrodynamic Interaction between Long and Thin Rods}

In order to calculate sedimentation velocities, the connection between
translational and angular velocities, and hydrodynamic forces and torques must be
found. In the present section such a relation will be established for two rods
on the Rodne-Prager level, that is, with the neglect of reflection contributions
between the rods. Considering only two rods limits the discussion
on the sedimentation velocity to first order in concentration.
Reflection contributions to the
two-rod hydrodynamic interaction functions and multi-body rod interactions
are both probably small in
comparison to the Rodne-Prager terms, due to the fact that the distance between
segments of different rods is of the order of the length of the rods, at least
in the isotropic state. A Rodne-Prager approximation could therefore work
quite well for long and thin rods, although explicit results for reflection
contributions should be obtained to confirm this intuition.

For the low Reynolds numbers under consideration, the translational
velocities $\vv_{j}$, $j=1,2$, and the angular velocities $\mathbf{\Omega}_{j}$
are linearly related to the hydrodynamic forces $\F^{h}_{j}$ and torques
$\T^{h}_{j}$ that the fluid exerts on the rods,
\begin{eqnarray}
\left( \begin{array}{c}
\vv_{1} \\
\vv_{2} \\
\mathbf{\Omega}_{1} \\
\mathbf{\Omega}_{2}
\end{array} \right) = \:-\;
\left( \begin{array}{cccc}
\mathbf{M}^{TT}_{11} & \mathbf{M}^{TT}_{12} & \mathbf{M}^{TR}_{11} & \mathbf{M}^{TR}_{12} \\
\mathbf{M}^{TT}_{21} & \mathbf{M}^{TT}_{22} & \mathbf{M}^{TR}_{21} & \mathbf{M}^{TR}_{22} \\
\mathbf{M}^{RT}_{11} & \mathbf{M}^{RT}_{12} & \mathbf{M}^{RR}_{11} & \mathbf{M}^{RR}_{12} \\
\mathbf{M}^{RT}_{21} & \mathbf{M}^{RT}_{22} & \mathbf{M}^{RR}_{21} & \mathbf{M}^{RR}_{22}
\end{array} \right) \cdot \left(
\begin{array}{c}
\F^{h}_{1} \\
\F^{h}_{2} \\
\T^{h}_{1} \\
\T^{h}_{2}
\end{array} \right) .
\end{eqnarray}
The superscripts ``T'' and ``R'' refer to translation and rotation,
respectively, while the superscript ``h'' on the forces and torques
refer to their hydrodynamic origin. On the Brownian time scale, the
$3\times 3$-dimensional mobility matrices $\mathbf{M}$ are
functions of the positions of the centers of the two rods and their
orientation.

As will turn out, in order to find the sedimentation velocity, we
need expressions for $\mathbf{M}_{1j}^{TT}$, for which
approximations are obtained in subsection~\ref{MM}. As a first
step, the fluid flow field generated by a translating rod must be
calculated. This is the subject of subsection~\ref{Flowfield}.
Rotation of rods also plays a role in sedimentation, but as will
turn out, to first order in concentration and with the neglect of
hydrodynamic reflection contributions, these do not contribute to
the sedimentation velocity. Explicit expressions pertaining to the
hydrodynamics of rotating rods are derived in the same spirit as
for translating rods in appendix A. Subsection~\ref{conclusion}
contains some concluding remarks.

For the hydrodynamic calculations the rods will be thought of as a rigid string of
spherical beads with diameter $D$. The length of the rods is $L$, and there are
$n+1=L/D$ beads per rod, with $n$ an even integer.

\subsection{Flow field generated by a translating rod}
\label{Flowfield}

The flow field generated by a rod that consists of $n+1$ beads is given by,
\begin{eqnarray}
\mathbf{u}(\r)\:=\:\sum_{j=-n/2}^{n/2}\,\oint_{\partial V_{j}}
dS' \,\T (\r-\r')\cdot\mathbf{f}_{j}(\r')\:,
\end{eqnarray}
where $\mathbf{T}$ is the Oseen tensor,
\begin{eqnarray}
\T (\r)\:=\:\frac{1}{8\pi \eta_{0} r} \left[ \hat{\mathbf{I}}+\hat{\r}\hat{\r} \right]\:,
\end{eqnarray}
with $\eta_{0}$ the shear viscosity of the solvent and $\hat{\r}=\r/r$ the unit
position vector. Furthermore, $\mathbf{f}_{j}(\r')$ is the force per unit area that
a surface element at $\r'$ of bead $j$ exerts on the fluid, and $\partial V_{j}$
is the spherical surface of bead $j$. For long and thin rods, the distances $\r$
of interest, relative to the positions of the beads, are those for which $r \gg D$,
with $D$ the diameter of the beads. Now write $\r'=\r_{j}+\mathbf{R}'$ with $\r_{j}$
the position coordinate of the $j^{th}$ bead, so that $R'=D/2$, and Taylor expand
the Oseen tensor in eq.(2) with respect to $\mathbf{R}'$. Keeping only the first
term in this Taylor expansion leads to relative errors of the order $R'/r \sim D/L$.
Up to that order we then find,
\begin{eqnarray}
\mathbf{u}(\r)\:=\:-\,\sum_{j=-n/2}^{n/2}\,\T(\r-\r_{j})\cdot \F_{j}^{h}\:,
\end{eqnarray}
with,
\begin{eqnarray}
\F_{j}^{h}\:=\:-\oint dS' \,\f_{j}(\r')\:,
\end{eqnarray}
the total force that the fluid exerts on bead $j$. With the neglect of end-effects
this force is equal for each bead, $\F_{j}\equiv \F^{h}/(n+1)= \frac{D}{L}\F^{h}$,
with $\F^{h}$ the total force on the rod.  Eq.(4) thus reduces to,
\begin{eqnarray}
\mathbf{u}(\r)\:=\:-\,\frac{D}{L}\sum_{j=-n/2}^{n/2}\,\T(\r-\r_{j})\cdot \F^{h}\:.
\end{eqnarray}
The force $\F^{h}$ is calculated in terms of the translational velocity of the
rod self-consistently from eq.(6) using Fax\'{e}n's theorem for translational motion
for each spherical bead, where the velocity $\mathbf{v}_{j}$ of bead $j$ is
expressed in terms of the force $\F_{j}^{h}$ on bead $j$ and the velocity $\mathbf{u}_{0}(\r_{j})$
at the center of the bead that would have existed without that bead being present,
\begin{eqnarray}
\mathbf{v}_{j}\:=-\frac{1}{3\pi \eta_{0} D}\F_{j}^{h}+\mathbf{u}_{0}(\r_{j})
+\frac{1}{24}D^{2}\nabla_{j}^{2} \mathbf{u}_{0}(\r_{j})\:,
\end{eqnarray}
where $\nabla_{j}$ is the gradient operator with respect to $\r_{j}$. The first term
on the right hand-side is just Stokes friction of a single bead in an unbounded
fluid, while the second term accounts for hydrodynamic interaction between
the beads. The fluid flow field $\mathbf{u}_{0}$ in turn is equal to,
\begin{eqnarray}
\mathbf{u}_{0}(\r)\:=\:\sum_{i=-n/2\:,\:i\neq j}^{n/2} \oint_{\partial V_{i}}
dS' \, \T(\r-\r')\cdot \f_{i}^{\ast}(\r')\:,
\end{eqnarray}
where $\f_{i}^{\ast}$ is the force per unit area that a surface element of bead
$i$ exerts on the fluid {\sl in the absence of bead $j$}. For very long rods,
consisting of many beads, the difference between $\f_{i}$ (the corresponding
force for the intact rod) and $\f^{\ast}_{i}$ may be neglected: there are
only a few neighbouring beads for which the difference is
significant, but there are many more beads further away from bead $j$ for which
the difference is insignificant. To within the same approximations involved
to arrive at eq.(4), eq.(8) can then be written as,
\begin{eqnarray}
\mathbf{u}_{0}(\r_{j})\:=\:-\sum_{i=-n/2\:,\:i\neq j}^{n/2}\,\T(\r_{j}-\r_{i})
\cdot \F_{i}^{h}\:.
\end{eqnarray}
Substitution of this expression into Fax\'{e}n's theorem (7), and using that
$\r_{j}-\r_{i}=(j-i)D\hat{\uu}$, with $\hat{\uu}$ the orientation of
the rod, leads to,
\begin{eqnarray}
\mathbf{v}_{j}&=&-\frac{1}{3\pi \eta_{0} D} \F_{j}^{h}
-\frac{1}{8\pi \eta_{0} D} \hat{\uu }\hat{\uu } \cdot \sum_{i=-n/2\:,\:i\neq j}^{n/2}
\left[ \frac{2}{\mid\! i-j \!\mid}- \frac{1}{6\mid\! i-j \! \mid^{3}} \right]
\cdot \F_{i}^{h} \nonumber \\
&&-\:\frac{1}{8\pi \eta_{0} D}\left[ \hat{\mathbf{I}}-\hat{\uu }\hat{\uu }\right]
\cdot \sum_{i=-n/2\:,\:i\neq j}^{n/2}
\left[ \frac{1}{\mid\! i-j \!\mid}+ \frac{1}{12\mid\! i-j \! \mid^{3}} \right]
\cdot \F_{i}^{h}\:,
\end{eqnarray}
where eq.(3) has been used, together with,
\begin{eqnarray}
\nabla^{2} \T(\r)\:=\: \frac{1}{4\pi \eta_{0} r^{3}} \left[ \hat{\mathbf{I}}
-3\hat{\r}\hat{\r} \right] \:.
\end{eqnarray}
For pure translational motion, the velocity $\mathbf{v}_{j}$ of each
bead is equal to the velocity $\mathbf{v}$ of the rod, so that both sides of
eq.(10) can be summed over $j$, yielding for the left hand-side $\vv\,L/D$.
Neglecting end-effects
and replacing sums by integrals (which is allowed for long and thin rods),
it is found that,
\begin{eqnarray}
\mathbf{v}\:=\:- \frac{\ln\{L/D\}}{4\pi \eta_{0} L} \left[ \hat{\mathbf{I}}
+\hat{\uu }\hat{\uu } \right] \cdot \mathbf{F}^{h}\:.
\end{eqnarray}
Notice that the Stokes friction contribution (the first term on the right
hand-side in eq.(10)) is logarithmically small in comparison to the friction
contribution due to hydrodynamic interaction between the beads. In fact, the
Stokes contribution is neglected in eq.(12). A matrix inversion, in order to
express $\F^{h}$ in terms of $\mathbf{v}$, and subsequent substitution into eq.(6),
after rewriting the sum over beads
as an integral over the center line of the rod, yields,
\begin{eqnarray}
\uu (\r )\:=\:\frac{4\pi \eta_{0}}{\ln\{ L/D\} }
\int_{-L/2}^{L/2} dl \;\T(\r-\r_{p}-l\hat{\uu })\cdot
\left[ \hat{\mathbf{I}}-\frac{1}{2}\hat{\uu }\hat{\uu } \right] \cdot \mathbf{v}\:,
\end{eqnarray}
with $\r_{p}$ the position coordinate of the rod.
This is the approximate expression for the fluid flow generated by a translating
long and thin rod that will be used in the following subsection to obtain an expression
for the mobility matrices $\mathbf{M}^{TT}_{1j}$, $j=1,2$.

\subsection{Calculation of $\mathbf{M}^{TT}$}
\label{MM}

In order to calculate the velocity $\vv_{2}$ that rod $2$ acquires
in the flow field (13) generated by a translating rod $1$, one
should in principle perform a reflection calculation up to very
high order : the field generated by rod $1$ is scattered by each
bead of rod $2$ and subsequently reflected hence and forth between
the different beads within rod 2. Such a calculation is hardly
feasible analytically. Here, the field generated by rod $1$ that is
incident on rod $2$ is approximated by a constant fluid flow field
$\bar{\uu }$ equal to the average of the incident field over the
center line of rod $2$. This ``hydrodynamic mean-field
approximation'' is accurate for distances of the order $L$ or
larger, for which separations the incident field indeed becomes
equal to a constant. For smaller distances between the rods this
procedure provides a semi-quantitative approximation. Within this
approximation, the velocity of rod $2$ immediately follows from
eq.(12), with $\vv
=\vv_{2}-\bar{\uu}$, $\F^{h}=\F_{2}^{h}$, the total force of the
fluid on rod $2$, and $\hat{\uu }=\hat{\uu }_{2}$, the orientation
of rod $2$,
\begin{eqnarray}
\vv_{2}\:=\:\bar{\uu } - \frac{\ln\{L/D\}}{4\pi \eta_{0} L} \left[ \hat{\mathbf{I}}
+\hat{\uu }_{2} \hat{\uu }_{2} \right] \cdot \F_{2}^{h}\:.
\end{eqnarray}
The average incident flow field follows from eqs.(13) and (12), with $\vv =\vv_{1}$, the velocity
of rod $1$ and $\F^{h}=\F^{h}_{1}$, the force on rod $1$,
\begin{eqnarray}
\bar{\uu }&=&\frac{4\pi \eta_{0} L}{\ln\{L/D\}} \,\mathbf{A} \cdot
\left[ \hat{\mathbf{I}}-\frac{1}{2} \hat{\uu }_{1} \hat{\uu }_{1} \right] \cdot \vv_{1} \nonumber \\
&=&-\mathbf{A} \cdot
\left[ \hat{\mathbf{I}}-\frac{1}{2} \hat{\uu }_{1} \hat{\uu }_{1} \right] \cdot
\left[ \hat{\mathbf{I}}+\hat{\uu }_{1} \hat{\uu }_{1} \right] \cdot \F^{h}_{1} \nonumber \\
&=& -\,\mathbf{A} \cdot \F_{1}^{h} \:,
\end{eqnarray}
where,
\begin{equation}
\mathbf{A}\:=\:\frac{1}{L^{2}} \int_{-L/2}^{L/2} dl_{1} \int_{-L/2}^{L/2} dl_{2} \,
\T(\r_{21}+l_{2}\hat{\uu }_{2}-l_{1}\hat{\uu }_{1})\:,
\end{equation}
with $\r_{21}=\r_{2}-\r_{1}$ the distance between the centers of
the two rods. Notice that for distances $\r_{21}$ between the
centers of the rods larger than $L$, the matrix $\mathbf{A}$
asymptotes to $\T(\r_{21})$. By definition the following
``mean-field'' expressions for the translational mobility matrices
are thus obtained (after an interchange of the indices $1$ and
$2$),
\begin{eqnarray}
\mathbf{M}^{TT}_{11}&=&\frac{\ln\{L/D\}}{4\pi \eta_{0}L}\left[ \hat{\mathbf{I}}
+\hat{\uu }_{1}\hat{\uu }_{1} \right] \:,\\
\mathbf{M}^{TT}_{12}&=&\frac{1}{L^{2}} \int_{-L/2}^{L/2} dl_{1}
\int_{-L/2}^{L/2}dl_{2}\,\T(\r_{12}+l_{1}\hat{\uu }_{1}-l_{2} \hat{\uu}_{2})\:.
\end{eqnarray}
One might try to device approximate expressions for the matrix $\mathbf{A}$.
However, sedimentation velocities are obtained as ensemble averages, also
with respect to orientations, giving rise to integrals with respect to $\r_{12}$ and
$\hat{\uu }_{1,2}$, which can be evaluated by numerical integration.

\subsection{Concluding Remarks}
\label{conclusion}

The approximations involved in the above discussion are justified
for very long rods, since $\mathcal{O}$(1)-constants are neglected
against terms of order $\ln\{ L/D\}$, both by neglecting
end-effects and replacing sums over beads by integrals (for the
evaluation of the sums in Fax\'{e}n's theorem in eq.(10)). Such
approximations are most important for the diagonal mobility matrix
$\mathbf{M}_{11}^{TT}$ (notice that factors $\ln\{L/D\}$ do not
appear in the off-diagonal matrice $\mathbf{M}_{12}^{TT}$, due to
the resubstitution of velocities in terms of forces). Both
end-effects and the mathematical approximations involved in the
calculation of the diagonal mobility matrix $\mathbf{M}_{11}^{TT}$
in eq.(17) may be accurately accounted for by the replacement,
\begin{eqnarray}
\ln\{L/D\}\;\;\rightarrow \;\; \ln\{L/D\}-\nu\:,
\end{eqnarray}
with $\nu = \nu_{\perp}$ and $\nu =\nu_{\parallel}$ a constant,
pertaining to translational motion perpendicular and parallel to
the rods orientation, respectively. This correction is
experimentally significant for somewhat shorter rods ($L/D < 20$,
say), but vanishes relatively to the logarithmic term for very long
rods. The actual values of $\nu_{\perp}$ and $\nu_{\parallel}$ for
cylindrical rods are equal to [4],
\begin{eqnarray}
\nu_{\perp}&=& -0.84 , \\
\nu_{\parallel}&=& 0.21 .
\end{eqnarray}
A more accurate expression for $\mathbf{M}_{11}^{TT}$ than in eq.(17) is,
\begin{eqnarray}
\mathbf{M}^{TT}_{11}\,=\,\frac{\ln\{L/D\}}{4\pi \eta_{0}L}\left[ \hat{\mathbf{I}}
+\hat{\uu }_{1}\hat{\uu }_{1} \right]\,-\,\frac{1}{4\pi\eta_{0} L}
\left[ \nu_{\perp} \hat{\mathbf{I}}+(2\nu_{\parallel}-\nu_{\perp})
\hat{\uu}_{1}\hat{\uu}_{1} \right] \,.
\end{eqnarray}
In the sequel we will use this expression for the mobility matrix
$\mathbf{M}_{11}^{TT}$ instead of eq.(17). The approximations
involved in the off-diagonal mobility matrices in eqs.(34,38) are
primarily due to the mean-field treatment of the incident flow
field. It is probably a formidable task to improve on these
expressions.

The use of the more accurate expression (22) for the diagonal translational
mobility matrix also circumvents the practical problem of calculating volume
fractions of colloidal rod material from given values for $L$, $D$ and
number density. For the bead model it is not so clear how the volume of a rod
must be expressed in terms of $L$ and $D$.
The volume of the cylindrical rod is simply equal to $\frac{\pi}{4}D^{2}L$.

\section{An Expression for the Sedimentation Velocity of Rods}

The sedimentation of colloidal material induces, through the
presence of the walls of the container, backflow of solvent. The
backflow velocity is inhomogeneous, and varies on the length scale
of the container. On a local scale, however, the backflow may be
considered homogeneous, and the sedimentation velocity can be
calculated relative to the local backflow velocity. This {\sl
relative} sedimentation velocity is a constant throughout the
container (except possibly in a small region of extent $L$ near the
walls of the container, where gradients of the backflow velocity
are large), and depends only on the properties of the suspension. A
formal evaluation of the sedimentation velocity directly from
eq.(1), by ensemble averaging, leads to spurious divergences, which
are the result of the neglect of the hydrodynamic effects of the
walls of the container which lead to solvent backflow. Batchelor
was the first to deal with these divergences correctly, and we will
use his arguments here [2].

Ensemble averaging of $\vv_{1}$ in eq.(1) gives the sedimentation
velocity $\vv_{s}$, which is thus found to be equal to,
\begin{eqnarray}
\vv_{s}\:=\:-<\mathbf{M}_{11}^{TT}\cdot \F_{1}^{h}
+\bar{\rho}\,V\mathbf{M}^{TT}_{12}\cdot \F_{2}^{h}
+\mathbf{M}^{TR}_{11}\cdot \T_{1}^{h}
+\bar{\rho}\,V\mathbf{M}^{TR}_{12}\cdot \T_{2}^{h}>\:,
\end{eqnarray}
where the brackets $<\cdots>$ denote ensemble averaging with respect to positions
and orientations of the rods. The factors $\bar{\rho}V=N\approx N\!-\!1$ account for the presence
of $N\!-\!1$ rods which all interact with rod $1$ under consideration. The divergence
problems mentioned above arising in the explicit evaluation of the ensemble averages will be dealt
with later.

In order to be able to calculate these ensemble averages, the forces and torques
must be expressed in terms of the positions and orientations of the rods. On the
Brownian time scale there is a balance between all the forces and torques on
each of the rods, that is, the total force and torque are equal to zero. The total
force in turn is equal to the sum of the force $\F^{h}_{j}$ that the fluid
exerts on the rod, the interaction force $\F^{I}_{j}=-\nabla_{j}\Phi$ (with
$\Phi$ the total interaction energy of the rods), the Brownian force
$\F^{Br}_{j}=-k_{B}T \nabla_{j}\ln\{ P\}$ (with $k_{B}$ Boltzmann's constant, $T$
the temperature and $P$ the probability density function for positions and orientations),
and the external force $\F^{ext}$ due to the gravitational field. Hence,
\begin{eqnarray}
\F_{j}^{h}\,=\,\nabla_{j} \Phi \,+\, k_{B}T\nabla_{j} \ln\{ P\}
\,-\,\F^{ext}\:.
\end{eqnarray}
Similarly, the total torque is the sum of the hydrodynamic torque $\T_{j}^{h}$,
the interaction torque $-\hat{\mathcal{R}}_{j}\Phi$, the Brownian torque
$-k_{B}T \hat{\mathcal{R}}_{j} \ln \{ P\}$, while the torque on each rod due to
the homogeneous external force vanishes. Hence,
\begin{eqnarray}
\T_{j}^{h}\,=\,\hat{\mathcal{R}}_{j} \Phi \,+\,k_{B}T \hat{\mathcal{R}}_{j}\ln \{ P \}\:,
\end{eqnarray}
where the rotation operator is defined as,
\begin{eqnarray}
\hat{\mathcal{R}}_{j}(\cdots )\:\equiv \:\hat{\uu}_{j} \times \nabla_{u_{j}} (\cdots )\:,
\end{eqnarray}
with $\nabla_{u_{j}}$ the gradient operator with respect to $\hat{\uu}_{j}$. Substitution
of eqs.(24,25) into eq.(23) for the sedimentation velocity yields,
\begin{eqnarray}
\mathbf{v}_{s}&=&-<\mathbf{M}_{11}^{TT} \cdot \left[
\nabla_{1} \Phi + k_{B}T\nabla_{1} \ln\{ P\} -\F^{ext} \right] \nonumber \\
&&\;\;\;\;\;\;\;\;\;\;\;\;\;\;\;\;\;\;\;\;
\bar{\rho}\,V\mathbf{M}_{12}^{TT}\cdot \left[ \nabla_{2} \Phi + k_{B}T\nabla_{2} \ln\{ P\}
-\F^{ext} \right]  \\
&&+\mathbf{M}_{11}^{TR}\cdot \left[
\hat{\mathcal{R}}_{1} \Phi +k_{B}T \hat{\mathcal{R}}_{1}\ln \{ P \}\right]
+\bar{\rho}\,V\mathbf{M}_{12}^{TR} \cdot \left[ \hat{\mathcal{R}}_{2} \Phi +k_{B}T \hat{\mathcal{R}}_{2}\ln \{ P \}\right] >\,. \nonumber
\end{eqnarray}

The next step in the explicit evaluation of these ensemble averages is to determine
the stationary probability density function $P\equiv P(\r_{1},\r_{2},\hat{\uu}_{1},\hat{\uu}_{2})$
for the positions and orientations of two rods. At this point it is convenient
to introduce the pair-correlation function $g$, defined as,
\begin{eqnarray}
P(\r_{1},\r_{2},\hat{\uu}_{1},\hat{\uu}_{2})\:\equiv \:
P(\r_{1},\hat{\uu}_{1})\,P(\r_{2},\hat{\uu}_{2})\,g(\r_{12},\hat{\uu}_{1},\hat{\uu}_{2})\:,
\end{eqnarray}
where $P(\r_{j},\hat{\uu}_{j})$ is the probability density function
for the position and orientation of a single rod. For spherical
particles in a homogeneous external gravitational field, the
probability density function for the position coordinates is simply
the equilibrium function, without an external field, provided that
the particles are identical. The probability density function
differs from the equilibrium function only in case the relative
sedimentation velocity of two spheres is different, for example due
to differing masses and/or sizes. For rods, things are somewhat
more complicated. Even if two rods are identical their relative
sedimentation velocity generally differs as a result of the fact
that the translational friction constant of rods is orientation
dependent (see eq.(17)). The probability density function is
generally dependent on the external force due to the fact that rods
with different orientations overtake each other during
sedimentation. Suppose, however, that the sedimentation velocity is
so slow, that during a relative displacement of two rods in the
gravitational field of the order of the length $L$ of the rods,
each rod rotated many times due to their Brownian motion. The
relative sedimentation velocity of two rods then averages out to
zero. For such a case, the pair-correlation function is only weakly
perturbed by the external field, so that we may use its equilibrium
Boltzmann form,
\begin{eqnarray}
g(\r_{1},\r_{2},\uu_{1},\uu_{2})\:=\: \exp\{ -\beta \Phi(\r_{1},\r_{2},\hat{\uu}_{1},\hat{\uu}_{2})\}\:,
\end{eqnarray}
Let us derive the inequality
that should be satisfied for eq.(29) to be valid. According to eq.(17), the largest
relative sedimentation velocity $\Delta \vv_{s}$ of two rods is approximately equal to,
\begin{eqnarray}
\mid\!\Delta \vv_{s}\!\mid \:\approx \: \frac{\ln \{ L/D\}}{4\pi \eta_{0}L} \mid\!\F^{ext}\!\mid\:.
\end{eqnarray}
On the other hand, the time $\tau_{rot}$ required for a rotational revolution
is equal to,
\begin{eqnarray}
\tau_{rot}\:=\:\frac{\pi\eta_{0}L^{3}}{3\,k_{B}T\,\ln\{L/D\}}\:.
\end{eqnarray}
The condition under which eq.(29) for the pair-correlation function is a
good approximation is therefore,
\begin{eqnarray}
\frac{L/\mid\!\Delta \vv_{s}\!\mid}{\tau_{rot}}\:\approx \:
\frac{12\,k_{B}T}{\mid\!\F^{ext}\!\mid \:L}\gg \:1\:.
\end{eqnarray}
Hence, the work provided by the external force to displace a
colloidal rod over a distance equal to its length should be much
smaller than a few times the thermal energy of the rods.
Substitution of typical numbers shows that this inequality is
satisfied under normal practical circumstances.\footnote{ For
example, for rods with a length of 100 nm, the sedimentation
velocity should be much less than 1 mm/min, in order that the
inequality (32) is satisfied.} Furthermore, when alignment of the
rods during sedimentation in a homogeneous suspension is of no
importance, the one-particle probability density functions in
eq.(28) are both constants equal to,
\begin{eqnarray}
P(\r_{j},\hat{\uu}_{j})\:=\: \frac{1}{4\pi \,V}\:.
\end{eqnarray}
We will restrict
ourselves here to the most common situation where the inequality (32) is satisfied,
and assume negligible alignment,
so that the probability density function is well approximated by
eqs.(28,29,33). In that case many of the terms in eq.(27) for
the sedimentation velocity cancel : the interaction contributions
cancel against the Brownian terms. The sedimentation velocity reduces simply to,
\begin{eqnarray}
\mathbf{v}_{s}&=&\left[ <\mathbf{M}_{11}^{TT}>
 +\,\bar{\rho}\,V<\mathbf{M}_{12}^{TT}> \right] \cdot\F^{ext} \:.
\end{eqnarray}
Since $\mathbf{M}_{12}^{TT}(\r_{12},\hat{\uu}_{1},\hat{\uu}_{2})\sim 1/r_{12}$
for large distances, its ensemble average with respect to position coordinates
diverges. Such a spurious divergence is also found for spherical particles, and
is the result of the neglect of the
hydrodynamic effect of the walls of the container.
Batchelor [2] showed that a formally divergent quantity, which is unambiguously finite valued on
physical grounds, can be subtracted from the ensemble average, rendering a perfectly
well defined sedimentation velocity. This finite
valued quantity is formally divergent for the same reason that the ensemble
average of $\mathbf{M}_{12}^{TT}$ is divergent, and subtraction accounts for
the local hydrodynamic effects of the wall.\footnote{ To within the approximations
made here, valid for long and thin rods, we do not encounter a conditionally
divergent contribution due to terms $\sim 1/r_{12}^{3}$, as for spheres. Such
a divergence can be dealt with by noting that the ensemble average of the
deviatoric part of the stress tensor vanishes. We will
not go into the extension of Batchelor's arguments to rods to deal with this
conditionally divergence problem.}
Batchelor's argument is as follows. First define the velocity
$\mathbf{u}(\r \!\mid\! \r_{1},\cdots ,\r_{N},\hat{\uu}_{1},\cdots ,\hat{\uu}_{N})$
as the velocity at a point $\r$ (either in the fluid or inside a colloidal rod),
given the positions $\r_{1},\cdots \r_{N}$ and orientations
$\hat{\uu}_{1},\cdots ,\hat{\uu}_{N}$ of $N$ rods. In the laboratory reference
frame the net flux of material through a cross sectional area must be zero.
This means that the ensemble average of $\uu$ must be zero. Hence,
\begin{eqnarray}
\mathbf{0}&=&<\mathbf{u}(\r \!\mid\! \r_{1},\cdots ,\r_{N},\hat{\uu}_{1},\cdots ,\hat{\uu}_{N})> \\
&&\!\!\!\!\!\!\!\!\!\!\!\!\!
=\int d\r_{1} \cdots \int d\r_{N} \oint d\hat{\uu}_{1}\cdots \oint d\hat{\uu}_{N}\;
\mathbf{u}(\r \!\mid\! \r_{1},\cdots ,\r_{N},\hat{\uu}_{1},\cdots ,\hat{\uu}_{N})
\,P(\r_{1},,\cdots ,\r_{N},\hat{\uu}_{1},\cdots ,\hat{\uu}_{N})\,,\nonumber
\end{eqnarray}
where $P(\r_{1},\cdots ,\hat{\uu}_{N})$ is the probability density function for
$\{ \r_{1},\cdots ,\hat{\uu}_{N}\}$.
Formally, this ensemble average diverges for the same reason that the sedimentation
velocity diverges : the flow field $\uu$ varies like $1/r$ for large distances due to its
Oseen contribution.
The formally divergent expression (35), that must be zero for physical reasons,
is subtracted from eq.(34) for the sedimentation velocity to render this
expression convergent. The field
$\uu$ can generally be written as the sum of a two terms : a term
to which the field would be equal to in the absence of reflection
contributions plus a term that accounts for the reflection contributions. To
within our approach reflection contributions are
neglected so that only the former term survives here. Hence,
\begin{eqnarray}
\uu (\r \!\mid \!\r_{1},\cdots ,\r_{N},\hat{\uu}_{1},\cdots ,\hat{\uu}_{N})&=&
\sum_{j=1}^{N} \uu (\r -\r_{j}) \;\;,\;\;f\!or\;\r\;in\;the\;f\!luid\:,\nonumber \\
&=&\vv_{s}\;\;\;\;\;\;\;\;\;\;\;\;\;\;\;\;\;\;,\;\;f\!or\;\r\;in\;a\;core\:,
\end{eqnarray}
where the field $\uu (\r -\r_{j})$, for $\r$ in the fluid, is the
sum of the flow fields in eq.(13), (with $\vv$ replaced by
$\vv_{s}-\uu_{s}$), and eq.(66) in appendix A, (both considered as
a function of the relative distance to the $j^{th}$ rod under
consideration), that is,
\begin{eqnarray}
\uu (\r-\r_{j})\:=\:\uu_{T}(\r-\r_{j}) +\uu_{R}(\r-\r_{j})\:,
\end{eqnarray}
where $\uu_{T}(\r -\r_{j})$ is the fluid flow generated by a translating rod as given in
eq.(13),
\begin{eqnarray}
\uu_{T}(\r-\r_{j})\:=\:\frac{1}{L}
\int_{-L/2}^{L/2} dl_{j} \;\T(\r-\r_{j}-l_{j}\hat{\uu }_{j})\cdot
\F^{ext}\:,
\end{eqnarray}
where the inverse of eq.(12) is used, together with $\F^{h}=-\F^{ext}$,
and $\uu_{R}(\r-\r_{j})$ is the field generated by a rotating rod, which is
similarly given by eq.(66) in appendix A.
Operating on both sides of eq.(35) with
$\int d\r \oint d\hat{\uu} P(\r,\hat{\uu})$, where $P(\r,\hat{\uu})$ is the constant
specified in eq.(33), and subtraction of the resulting equation from
eq.(34) for the sedimentation velocity yields, for identical rods and to first order
in concentration (rename $\r_{1}=\r_{2},\,\r =\r_{1},\, \hat{\uu}_{1}=\hat{\uu}_{2}$ and
$\hat{\uu} =\hat{\uu}_{1}$),
\begin{eqnarray}
\vv_{s}&=&-\varphi\,\vv_{s}\, + \left[ <\mathbf{M}_{11}^{TT}>+\bar{\rho}\,V<\mathbf{M}_{12}^{TT}>
\right] \cdot \F^{ext} \nonumber \\
&&\;\;\;\;\;
-\,\frac{\bar{\rho}}{(4\pi )^{2}}\int d\r_{12} \oint d\hat{\uu}_{1} \oint d\hat{\uu}_{2}
\left[\uu_{T}(\r_{12})+\uu_{R}(\r_{12})\right]\,\chi_{f}(\r_{1}\!\mid\!\r_{2},\hat{\uu}_{2})\:,
\end{eqnarray}
where $\chi_{f}$ is the characteristic function that restricts the integrations to
points $\r$ which are in the fluid, not inside the core of rod $2$,
\begin{eqnarray}
\chi_{f} (\r\!\mid\!\r_{2},\hat{\uu}_{2})&=&1\;\;,\;\; f\!or\;\r\;in\;the\;f\!luid\;,\nonumber \\
&=&0\;\;,\;\; f\!or\;\r\;in\;the\;core\;of\;rod\;2\;.
\end{eqnarray}
Without interactions the angular velocity of each rod is simply proportional
to the hydrodynamic
torque on the same rod (see eqs.(59,64) in appendix A), which hydrodynamic torques are zero.
Since the integral in the above expression for the sedimentation velocity
is multiplied by the concentration $\bar{\rho}$, it follows that the rotational
field $\uu_{R}$ does not contribute to first order in the density. For the
same reason, in each term that is multiplied by the concentration,
$\F^{ext}$ may be expressed with eq.(22) in terms of the sedimentation velocity
$\vv_{s}^{0}$ without interactions, at infinite dilution, as,
\begin{eqnarray}
\vv_{s}^{0}\:=\:\left[ \frac{\ln\{L/D\}}{3\pi \eta_{0} L}
-\frac{1}{6\pi\eta_{0}L}(\nu_{\perp} + \nu_{\parallel})
\right]\,\F^{ext}\:.
\end{eqnarray}
We thus find the following expression for the sedimentation
velocity, valid to first order in volume fraction,
\begin{eqnarray}
\vv_{s}\:=\:\vv_{s}^{0} \,\left[ 1\,-\,\left(
\frac{f_{1}+f_{2}}{2 \ln\{ L/D\}-(\nu_{\perp}+\nu_{\parallel})}
+\mathcal{O}(D/L)
\right) \frac{L}{D}\,\varphi \right]\:,
\end{eqnarray}
where the functions $f_{1}$ and $f_{2}$ are  equal to,
\begin{eqnarray}
f_{1}&=&-\frac{1}{4\pi^{3}\, D L^{3}}\int d\r_{12} \oint d\hat{\uu}_{1}
\oint d\hat{\uu}_{2} \left[ g(\r_{12},\hat{\uu}_{1},\hat{\uu}_{2})-\chi_{f}(\r_{1}\!\mid\!\r_{2},\hat{\uu}_{2}) \right] \nonumber \\
&&\;\;\;\;\;\;\;\;\;\;\;\;\;\;\;\;\;\;\;\;\;\;\;\;\;\;\;\;\;\;\;\;\;\times\:
\int_{-L/2}^{L/2} dl_{1} \int_{-L/2}^{L/2} dl_{2}\,
\frac{1}{\mid\!\r_{12}+l_{1}\hat{\uu}_{1}-l_{2}\hat{\uu}_{2}\!\mid}\,, \\
f_{2}&=&-\frac{1}{4\pi^{3}\, D L^{3}}\int d\r_{12} \oint d\hat{\uu}_{1}
\oint d\hat{\uu}_{2} \,\chi_{f}(\r_{1}\!\mid\!\r_{2},\hat{\uu}_{2})\nonumber \\
&&\;\;\;\;\;\;\times \:
\int_{-L/2}^{L/2} dl_{1} \int_{-L/2}^{L/2} dl_{2}\,
\left[
\frac{1}{\mid\!\r_{12}+l_{1}\hat{\uu}_{1}-l_{2}\hat{\uu}_{2}\!\mid}
\,-\,\frac{1}{\mid\!\r_{12}-l_{2}\hat{\uu}_{2}\!\mid} \right] \,.
\end{eqnarray}
where the expressions (13) and (38) for the mobility matrix $\mathbf{M}_{12}^{TT}$
and the field $\uu_{T}$ are used, respectively.
We also used that integrals
over the Oseen tensor must be proportional to the identity tensor, so that
in these integrals the Oseen tensor may be replaced by the trace $\frac{1}{3}Tr\{ \T \}$
of that tensor, which, according to its defining equation (3), is equal to,
\begin{eqnarray}
Tr \{ \T (\r) \}\:=\: \frac{1}{2 \pi \eta_{0} r}\:.
\end{eqnarray}
For rods interacting only via a hard-core repulsion, it is shown in appendix B
how to reduce the number of integrations, leading to the following results,
\begin{eqnarray}
f_{1}&=&\frac{8}{\pi^{3}}\int_{0}^{\infty} \!\!\!dx \int_{-1}^{1} dz_{1}
\int_{-1}^{1} dz_{2}\int_{0}^{\pi} d\Psi\, \nonumber \\
&&\!\!\!\!\!\!\!\!\!\!\!\!\!\!\times \,j_{0}^{2}(x\,z_{1})\:j_{0}^{2}(x\,z_{2})
\left[ 1\,- \left( z_{1}\,z_{2}
+\sqrt{(1-z_{1}^{2})(1-z_{2}^{2})}\:\cos\{\Psi\}\right)^{2}\right]^{1/2}
+\mathcal{O}(D/L)\, \nonumber \\
&&\;\;\;\;\;\;\;\;\;\;\;=\:6.4 \cdots \,+\,\mathcal{O}(D/L)\:, \\
f_{2}& = &\frac{2}{9}\,\frac{L}{D}\,+\,\mathcal{O}(D/L)\:.
\end{eqnarray}
The numerical value for $f_{1}$ has been obtained by numerical integration
and applies for hard-core interactions, where $g$ is equal to $0$ when two cores
overlap, and equal to $1$ otherwise. The result for $f_{2}$ is independent of
the kind of direct interaction between the rods.

Substitution of the numerical values for $f_{1}$ and $f_{2}$ from
eqs.(46,47) and into eq.(42) gives our final result for the
sedimentation velocity up to $\mathcal{O}(D/L)$ contributions,
\begin{eqnarray}
\vv_{s}\:=\:\vv_{s}^{0} \,\left[ 1\,-\,
\frac{6.4 \,+\,\frac{2}{9}\,\frac{L}{D}}{2 \ln\{ L/D\}-(\nu_{\perp}+\nu_{\parallel})}
 \frac{L}{D}\,\varphi \right]\:,
\end{eqnarray}
The volume fraction prefactor,
\begin{eqnarray}
\alpha \:=\:
\frac{6.4 \,+\,\frac{2}{9}\,\frac{L}{D}}{2 \ln\{ L/D\}-(\nu_{\perp}+\nu_{\parallel})}
 \frac{L}{D}\:,
\label{Dhont}
\end{eqnarray}
is plotted as a function of $L/D$ in Fig.1 (where both
$\nu_{\perp}$ and $\nu_{\parallel}$ are taken equal to $0$). Also
plotted is an older result due to Peterson [3] who predicted,
\begin{eqnarray}
\alpha\:=\:\frac{8 \,(3/8)^{2/3} \,(L/D)^{1/3}}{2\ln\{ L/D\}}\,\frac{L}{D}\:.
\label{Peterson}
\end{eqnarray}
In this latter theory back flow is not correctly accounted for,
hydrodynamic interactions are orientationally preaveraged and
certain integrals are not precisely calculated but only estimated.
The data points shown in Fig. 1a are experimental results for
silica rods, coated with stearyl alcohol and disolved in
cyclohexane. The data point $\times$ is taken from Ref. [5]. The
point $\circ$ is an unpublished result from the same author's. The
data point if Fig. 1b is a data point for {\it fd}-virus at high
salt concentration, as obtained in the experimental section of the
present paper. As can be seen from Fig. 1, the present prediction
is virtually equal to that of Peterson for $L/D < 30$, but large
differences are found for large aspect ratios. For large aspect
ratios, $\alpha$ is predicted to vary like $\sim
\left( L/D
\right)^{2}/\ln\{ L/D\}$, in contrast to Peterson's result $\sim
\left( L/D \right)^{4/3}/\ln
\{ L/D \}$. For smaller aspect ratios $\alpha$ approaches
approximately Batchelor's value for spheres $\alpha =6.55$, which
is probably fortuitous in view of the approximations made here
which limit the results to be meaningful only for long and thin
rods.

\section{Experimental Results}

The concentration dependence of the sedimentation velocity
predicted by Eq. (\ref{Dhont}) differs significantly  from
Peterson's result Eq. (\ref{Peterson}) only for rods with large
L/D. In our experiments we have used filamentous bacteriophage {\it
fd} which is a rod-like virus with L/D $\approx$ 130. Other
relevant physical characteristics of {\it fd}, are it's length
$L=880$nm, it's diameter $D= 6.6$ nm [6], and it's density of 1.285
mg/ml [7]. Because of it's large L/D ratio the virus is a
semi-flexible rather then a rigid rod characterized with
persistence length of 2.2 $\mu$m [8]. It's linear charge density is
10 e$^{-}$/nm at pH 8.2 [9].

We have grown {\it fd} virus according to standard procedures of
molecular biology described in Maniatis [10]. The virus suspension
was first purified in a cesium chloride density gradient and then
extensively dialyzed against tris buffer at pH 8.15 and at the
desired ionic strength for the sedimentation experiments. After
that the virus was concentrated by ultracentrifugation and from
this stock solution a series of samples with different
concentrations were prepared. The sedimentation velocity was
measured on a Beckman XL-A analytical ultracentrifuge equipped with
UV absorbance optics. Most of the experiments were done at 25
$^{\circ}$C and at a centrifugal force equal to 45,500 g's (25,000
rpm). Before each sedimentation experiment the sample and rotor
were allowed to equilibrate at the desired temperature for a few
hours. Sedimentation data showed some unexpected features,
interfering with straightforward calculation of the sedimentation
coefficient. For this reason we have added appendix C, where the
detailed analysis of our data is given.

The measured sedimentation velocity for a range of volume fractions
of {\it fd} from dilute solution up to a stable nematic phase are
shown in Fig. \ref{10mMSvedberg}. All the samples in these
measurements were kept at 8mM ionic strength. The sedimentation
velocity of rods in the isotropic phase uniformly decreases with
increasing concentration. After Ref. [10] we have tried to fit our
experimental data to a functional form $S_{\phi}=S_0(1-p\phi)^\nu$
where $S_0$ is the sedimentation velocity at infinite dilution and
$\phi$ is the volume fraction of rods. The experimental values of
sedimentation velocity are reported in Svedbergs where 1 $S =
10^{-13} s^{-1}$. As seen from Fig. \ref{10mMSvedberg} we obtain a
reasonable fit to the experimental data in the isotropic phase and
find that the sedimentation velocity at infinite dilution is
$S_0=46.0$ for the value of constants $\nu=-1/3$ and p=3600. After
linearizing our fitted formula we find that the volume prefactor
$\alpha
\approx 1200$ is much larger then predicted by Eq.
(\ref{Dhont}). The reason for such a high value of slope $\alpha$
is the low ionic strength at which the experiment was performed.
The same increase in $\alpha$ with decreasing ionic strength is
observed in sedimentation of spherical particles [12,13]. Also we
note that in this case the region where the sedimentation velocity
varies linearly with rod concentration is limited to very low
volume fraction of rods.

It is a well known fact that elongated rods at high volume
fractions undergo a first order phase transition to a liquid
crystalline nematic phase [14]. The nematic phase is characterized
by a short range liquid-like positional order and long range
solid-like orientational order of rods. {\it fd} virus forms  a
cholesteric phase instead of the nematic phase [6,15]. Locally the
cholesteric phase is equivalent to nematic, however on a
macroscopic scale the average direction of molecules in a
cholesteric phase forms a helix. The free energy difference between
a cholesteric phase and a nematic phase is very small and although
our experiments are performed on the cholestric phase only, we
expect that our results are generic and would hold for a nematic
phase of hard rods as well.

Bacteriophage {\it fd} in the cholesteric phase exhibits
qualitatively new behavior when placed in centrifugal field.
Instead of a single sedimenting boundary and single plateau we
observe two boundaries with two plateau's sedimenting at different
velocities  as shown in Fig. \ref{Time_Series_Nematic}. To confirm
that this change in sedimentation behavior is indeed due to the
formation of the nematic phase we have made a sample which is
co-existing between the isotropic and nematic phase. After the
sample had phase separated into macroscopically distinct
co-existing phases, each phase was sedimented separately. In the
isotropic phase there was no sign of a second  boundary, while in
the nematic (cholesteric) phase we observed a fast sedimenting
second boundary that was slightly more concentrated then the first
component. On one hand, the slow component had a plateau
concentration and a sedimentation velocity that was almost
independent of the average concentration. On the other hand, the
sedimentation velocity of the faster moving component rapidly
decreases with increasing $\it fd $ concentration and at the same
time the difference between the plateau concentrations of the two
components increases with increasing average concentration.

The unstable sedimentation of colloidal rods in the nematic phase
has a similar origin as the self-sharpening effect described in
Appendix C. The reason for the instability is the discontinous jump
in sedimentation velocity that occurs at the isotropic-nematic
phase transition as is shown in Fig. \ref{10mMSvedberg} [16]. The
denser nematic phase sediments at a significantly higher velocity
then a more dilute isotropic phase. Initially a stable nematic
phase occupies the whole sample length. When the centrifugal field
is turned on a sharp sedimenting boundary starts moving towards the
bottom of the container. Below this boundary (to the right) the
rods are still in the nematic phase while above it the
concentration of rods is very low and therefore they are in the
isotropic state. This occurs because some particles will inevitably
diffuse against the centrifugal field from a highly concentrated
plateau into the dilute region. As this happens they simultaneously
undergo a transition from the nematic to the isotropic phase. Since
the sedimentation velocity of rods in the isotropic phase is much
lower then in nematic phase, the probability of the rods diffusing
from the isotropic phase back into the nematic phase is virtually
zero. It is this asymmetry that results in a continuous flux of
particles from the nematic into the isotropic phase and contributes
to the formation of the second plateau in the isotropic phase that
is moving at a slower speed. The concentration of the rods in the
isotropic plateau will be very close to the concentration of
isotropic rods co-existing with the nematic phase due to the
self-sharpening effect because dilute isotropic rods will catch up
with more concentrated isotropic rods. However, the highest
concentration the isotropic rods can attain is the co-existence
concentration between the isotropic and nematic phases, as long as
the nematic phase sediments faster then the isotropic phase.
Indeed, this is very close to what we observe in Fig.
\ref{Time_Series_Nematic}. Another experimental observation
corroborating our explanation is that the sedimentation velocity
and concentration of the slower isotropic plateau does not change
significantly with average concentration of rods as seen in Fig
\ref{10mMSvedberg}.

 Since the theory presented in this paper is valid only to first
order in concentration of rods, to obtain an accurate value of the
prefactor $\alpha$  in Eq. \ref{Dhont} we have made additional
measurements in the dilute to semi-dilute range. Our results are
presented in Fig. \ref{Svedberg}. We note that the overlap to
semi-dilute concentration for {\it fd} with $L=0.88 \mu$m is at
volume fraction of $5.9\cdot 10^{-5}$. Unlike the previous
measurements, we have done these measurements at high ionic
strength where the behavior of charged rods is expected to approach
the behavior of hard rods. Additionally at high ionic strength we
expect the sedimentation velocity to have a linear dependence on
volume fraction of rods up to higher values of volume fraction. The
results for ionic strength of 50mM and 100mM ionic strength are
shown in figure \ref{Svedberg}. The volume prefactor in Eq.
(\ref{Dhont}) at 50 mM ionic strength is $\alpha = 450\pm40$ and at
100mM ionic strength $\alpha=440\pm60$ . We have repeated the
experiment at 100mM ionic strength on a different analytical
Beckman Xl-A ultracentrifuge and obtained the following result
$\alpha=490 \pm 50$. We conclude that $\alpha=470\pm50$ which is
the result plotted in Fig. 1b. Since the values of the coefficient
$\alpha$ do not change much with changing ionic strength from 50 mM
to 100mM we conclude that the charged rods have approached the hard
rod limit. Note that because of it's large L/D ratio $fd$ is
slightly flexible with a persistence length which is 2.5 times it's
contour length [7]. Still for the experimentally determined
parameters of $fd$, which are $L=880$nm and $D=6.6$nm, our
experimental results compare favorably to the Eq. \ref{Dhont},
which predicts the value of $\alpha=488$ (see Fig. 1). In contrast,
the previous result due to Peterson  in Eq. \ref{Peterson} predicts
a lower value of $\alpha=288$

\section{Acknowledgment}
We acknowledge valuable discussions with R. B. Meyer. This research
was supported by National Science Foundations grants No.
DMR-9705336, INT-9113312 and by the Netherlands Foundations of
Fundamental Research(FOM). Additional information is avaliable
online: www.elsie.brandeis.edu.

\section*{Appendix A}

{\bf Flow field generated by a rotating rod} \newline

Consider a rod with its center at the origin, which rotates with an angular velocity
$\mathbf{\Omega}$. The angular velocity is decomposed in its component perpendicular
and parallel to the rods center line,
\begin{eqnarray}
\mathbf{\Omega}_{\perp}&=&\left[ \hat{\mathbf{I}}-\hat{\uu}\hat{\uu}\right]
\cdot \mathbf{\Omega} \:,\\
\mathbf{\Omega}_{\parallel}&=&\hat{\uu}\hat{\uu}\cdot \mathbf{\Omega} \:.
\end{eqnarray}
Due to the linearity of the governing hydrodynamic equations, the flow fields
generated by a rod rotating along $\mathbf{\Omega}_{\perp}$ and $\mathbf{\Omega}_{\parallel}$
may be calculated separately and added to obtain the flow field of the rod rotating
along $\mathbf{\Omega}$.

Let us first consider a rod rotating with an angular velocity
$\mathbf{\Omega}_{\perp}$. The flow field that is generated by this
rotating rod is given by the general equation (2). The relative
change of the velocity of the beads is $\sim 1/j$. For beads
further away from the origin one may therefore consider the
velocity over larger groups of beads as being virtually constant.
The force on bead $j$ is then proportional to its own velocity,
\begin{eqnarray}
\F_{j}^{h}\:=\:-C\,\mathbf{\Omega}_{\perp}\times \r_{j}\:=\:
-C\,D\,j\,\mathbf{\Omega}_{\perp}\times \hat{\uu}\:.
\end{eqnarray}
where $C$ is an as yet unknown proportionality constant. This
expression is not valid for beads close to the center of the rod :
for these beads the forces may have a different direction than
their velocity. The fluid flow field generated by a long and thin
rod, however, is primarily determined by the relatively large
velocities of the beads further away from its center. We may
therefore use eq.(53), except for relatively few beads close to the
center and near the tips of the rod. Since $\r_{j}=jD\hat{\uu}$,
the torque is thus found, to leading order in $D/L$, to be equal
to,
\begin{eqnarray}
\T^{h}_{\perp}\:=\:\sum_{j=-n/2}^{n/2} \r_{j}\times \F_{j}^{h}\:=\:
-\,C\,D^{2}\,\frac{1}{12}\left( \frac{L}{D} \right)^{3} \,\hat{\uu}\times
(\mathbf{\Omega}_{\perp}\times \hat{\uu})\:=\:
-\,C\,D^{2}\,\frac{1}{12}\left( \frac{L}{D} \right)^{3}\mathbf{\Omega}_{\perp}\:,
\end{eqnarray}
since $\mathbf{\Omega}_{\perp}$ is perpendicular to $\hat{\uu}$.
It is used here that $\sum_{j=1}^{k} j^{2}= \frac{1}{6}k(k+1)(2k+1)$.
First of all, the constant $C$ is
calculated self-consistently from Fax\'{e}n's theorem in the form
of eq.(10). Multiplying both sides of eq.(10) by
$\r_{j}\times $,
using that $\r_{j}\times \vv_{j}=j^{2}D^{2}\mathbf{\Omega}_{\perp}$,
and summation over beads, leads to,
\begin{eqnarray}
\frac{1}{12}\left( \frac{L}{D} \right)^{3} D^{2} \mathbf{\Omega}_{\perp}\,=\,
-\frac{1}{3\pi \eta_{0} D}\T^{h}_{\perp} +\frac{C\,D}{8\pi \eta_{0}} \left( \frac{L}{D}
\right)^{3} \:g(L/D)\:\mathbf{\Omega}_{\perp}\:,
\end{eqnarray}
where the function $g$ is defined as,
\begin{eqnarray}
g(L/D)\:=\:\frac{1}{(n+1)^{3}} \sum_{j=-n/2}^{n/2}\;\;\sum_{i=-n/2\:,\:i\neq j}^{n/2}
ij\left[ \frac{1}{\mid\!i-j\!\mid}+\frac{1}{12}\frac{1}{\mid\!i-j\!\mid^{3}}\right]\:.
\end{eqnarray}
For long and thin rods the summations may be replaced by integrals, leading to,
\begin{eqnarray}
g(L/D)\:=\:\frac{1}{6}\,\ln\{ L/D \}\:,
\end{eqnarray}
up to leading order in $D/L$. Substitution of eq.(17) for the torque yields
a single equation for $C$, yielding, again up to leading order,
\begin{eqnarray}
C\:=\:\frac{4\pi \eta_{0} D}{\ln \{ L/D \}}\:.
\end{eqnarray}
Hence, from eq.(17),
\begin{eqnarray}
\mathbf{\Omega}_{\perp}\:=\:-\frac{3\ln\{ L/D\}}{\pi \eta_{0} L^{3}} \,\T^{h}_{\perp}\:.
\end{eqnarray}
The flow field $\uu_{\perp}$ that is generated by a rotating rod
may now be obtained from eq.(4), to within the same approximations
that were discussed in the section~\ref{Flowfield}, as,
\begin{eqnarray}
\uu_{\perp}(\r)\,=\,-\sum_{j=-n/2}^{n/2} \T (\r-\r_{j})\cdot \F_{j}^{h}\,=\,
\frac{4\pi \eta_{0} D^{2}}{\ln \{L/D\}} \sum_{j=-n/2}^{n/2}
\,\T(\r-\r_{j})\cdot (\mathbf{\Omega}_{\perp}\times j\,\hat{\uu})\,.
\end{eqnarray}
Replacing the sum over beads by a line integral, we thus find,
\begin{eqnarray}
\uu_{\perp}(\r)\,=\,\frac{4\pi \eta_{0}}{\ln \{L/D\}} \int_{-L/2}^{L/2} dl \:
\T(\r-\r_{p}-l\hat{\uu})\cdot (\mathbf{\Omega}_{\perp}\times l\,\hat{\uu})\,,
\end{eqnarray}
where $\r_{p}$ is the position coordinate of the rod.

Next consider a rod rotating with an angular velocity
$\mathbf{\Omega}_{\parallel}$. For this case we have to resort to
Fax\'{e}n's theorem for rotational motion of a bead, which reads,
\begin{eqnarray}
\mathbf{\Omega}_{\parallel}\:=\:-\,\frac{1}{\pi\eta_{0} D^{3}} \T_{j}^{h}
\,+\,\frac{1}{2}\,\nabla_{j}\times \mathbf{u}_{0}(\r_{j})\:,
\end{eqnarray}
where, as in the translational Fax\'{e}n's theorem (7), $\mathbf{u}_{0}$ is
the fluid flow velocity that would have existed in the absence of bead $j$. The first term
on the right hand-side is just Stokes rotational friction of a single bead in an unbounded
fluid, while the second term accounts for hydrodynamic interaction between the beads.
The important thing to note here is that the fluid flow generated by a single
rotating bead is now equal to,
\begin{eqnarray}
\uu_{j}(\r)\:=\:\left( \frac{D/2}{\mid\!\r -\r_{j}\!\mid} \right)^{3} \mathbf{\Omega}_{\parallel}\times (\r-\r_{j})\:,
\end{eqnarray}
so that this fluid flow is $\mathbf{0}$ along the entire center
line of the rod. This implies that hydrodynamic interaction between
the beads is unimportant for this case. For a long and thin rod
rotating along its center line, each bead experiences a rotational
friction that is practically equal to the Stokes friction, as if
each bead where alone in an unbounded fluid. As a result, the total
torque on the rod is simply the sum of the Stokesian torques on the
beads, so that it follows immediately from Fax\'{e}n's theorem (62)
that,
\begin{eqnarray}
\mathbf{\Omega}_{\parallel}\:=\:-\,\frac{1}{\pi \eta_{0} D^{2} L} \,\T^{h}_{\parallel}\:.
\end{eqnarray}
Furthermore, the total fluid flow $\uu_{\parallel}$ is simply the
sum of the fluid flows (63) generated by the rotating beads as if
they were alone in an unbounded fluid, since hydrodynamic
interaction between the beads is unimportant in the present case.
Replacing the sum by a line integral thus yields,
\begin{eqnarray}
\uu_{\parallel}(\r)\:=\:\frac{D^{2}}{8}\,\int_{-L/2}^{L/2}dl\,
\frac{1}{\mid\!\r -\r_{p}-l\,\hat{\uu}\!\mid^{3}}\,\left( \mathbf{\Omega}_{\parallel}\times
(\r -\r_{p})\right)\:.
\end{eqnarray}

The fluid flow $\uu =\uu_{\perp}+\uu_{\parallel}$ generated by a
rotating rod with an arbitrary angular velocity
$\mathbf{\Omega}=\mathbf{\Omega}_{\perp}+\mathbf{\Omega}_{\parallel}$
follows by combining eqs.(51,52) and (60,65),
\begin{equation}
\uu (\r)=
\frac{4\pi \eta_{0}}{\ln \{L/D\}} \int_{-L/2}^{L/2} dl \,
\T(\r-l\hat{\uu})\cdot (\mathbf{\Omega}\times l\,\hat{\uu})
+\frac{D^{2}}{8}\,\int_{-L/2}^{L/2}dl\,
\frac{1}{\mid\!\r -\r_{p}-l\,\hat{\uu}\!\mid^{3}}\,\left( (\hat{\uu}\hat{\uu}\cdot\mathbf{\Omega})\times
(\r -\r_{p}) \right) \:.
\end{equation}
This approximate expression will be used in the following paragraph to obtain an
expression for the mobility matrices $\mathbf{M}^{TR}_{1j}$, $j=1,2$.

\vspace{0.50cm}

{\bf Calculation of $\mathbf{M}^{TR}$} \newline

In order to calculate the velocity $\vv_{2}$ that rod $2$ acquires
in the flow field (66) generated by a rotating rod $1$, we apply,
without further discussion, the same ``mean-field'' approach as in
the previous section. The velocity $\vv_{2}$ is approximated by
taking the fluid flow field generated by the rotating rod as a
constant, equal to the average of the actual field over the center
line of the rod. Hence,
\begin{eqnarray}
\vv_{2}\:=\:\bar{\uu } - \frac{\ln\{L/D\}}{4\pi \eta_{0} L} \left[ \hat{\mathbf{I}}
+\hat{\uu }_{2} \hat{\uu }_{2} \right] \cdot \F_{2}^{h}\:,
\end{eqnarray}
where the average flow field in terms of the torque on rod $1$
follows from eqs.(66) and (59,64), with
$\mathbf{\Omega}=\mathbf{\Omega}_{1}$, the angular velocity of rod
$1$ and $\T_{1}^{h}$ the torque on rod 1,
\begin{eqnarray}
\bar{\uu }&=&\frac{12}{L^{4}}
\int_{-L/2}^{L/2} dl_{1}
\int_{-L/2}^{L/2}dl_{2}\,\T(\r_{21}+l_{2}\hat{\uu }_{2}-l_{1} \hat{\uu}_{1})
\cdot \left( l_{1}\,\hat{\uu}_{1}\times \T_{1}^{h} \right) \nonumber \\
&+&\frac{1}{8\pi \eta_{0} L^{2}}\int_{-L/2}^{L/2} dl_{1}
\int_{-L/2}^{L/2}dl_{2}\,\frac{1}{\mid\!\r_{21}+l_{2}\hat{\uu }_{2}-l_{1} \hat{\uu}_{1}
\!\mid^{3}}(\r_{21}+l_{2}\,\hat{\uu}_{2}) \times \left( \hat{\uu}_{1}\hat{\uu}_{1}
\cdot \T_{1}^{h} \right)\:.
\end{eqnarray}
By definition the following ``mean-field'' expression for the
translational-rotational mobility matrices are thus obtained (after
an interchange of the indices $1$ and $2$),
\footnote{ The outer product $\mathbf{a}\times
\mathbf{A}$ of a vector $\mathbf{a}$ and a matrix $\mathbf{A}$ is defined as
the matrix with column vectors equal to the outer product of $\mathbf{a}$ and
the column vectors of $\mathbf{A}$. The outer product is thus taken with
respect to the first index on $\mathbf{A}$.}
\begin{eqnarray}
\mathbf{M}_{11}^{TR}&=&\mathbf{0}\:,\\
\mathbf{M}_{12}^{TR}&=&\frac{12}{L^{4}}\int_{-L/2}^{L/2}dl_{1}
\int_{-L/2}^{L/2}dl_{2}\:l_{2}\,\hat{\uu}_{2}\times \T (\r_{12}+l_{1}\hat{\uu}_{1}-l_{2}
\hat{\uu}_{2}) \nonumber \\
&+&\frac{1}{8\pi \eta_{0} L^{2}} \int_{-L/2}^{L/2} dl_{1} \int_{-L/2}^{L/2}dl_{2}\:
\frac{1}{\mid\!\r_{12}+l_{1}\,\hat{\uu}_{1}-l_{2}\,\hat{\uu}_{2}\!\mid^{3}}
\left[ \hat{\uu}_{2}\times (\r_{12}+l_{1}\,\hat{\uu}_{1})\right] \hat{\uu}_{2}\:.
\end{eqnarray}
A non-zero contribution to $\mathbf{M}_{11}^{TR}$ stems entirely from reflection
contributions, since a pure rotation of a single rod in an unbounded fluid
does not induce a translational velocity of the same rod.
As mentioned before, reflection contributions
are small in the isotropic state, since the typical distance between the beads
of different rods is of the order $L$.

\section*{Appendix B}

As a first step in the evaluation of the integrals in eq.(43) for
$f_{1}$, the Fourier transform of the Oseen tensor ($\T
(\mathbf{k})=\frac{1}{\eta_{0}k^{2}}\left[ \hat{\mathbf{I}}-
\hat{\mathbf{k}}\hat{\mathbf{k}}\right]$, with
$\hat{\mathbf{k}}=\mathbf{k}/k$) is substituted, and the
integrations with respect to $l_{1}$ and $l_{2}$ are performed,
with the result,
\begin{eqnarray}
f_{1}&=&-\frac{1}{8\pi^{5}\, D L}\int d\mathbf{k}\,k^{-2}   \oint d\hat{\uu}_{1}
\oint d\hat{\uu}_{2} \int d\r_{12}\,\left[ g(\r_{12},\hat{\uu}_{1},\hat{\uu}_{2})-\chi_{f}(\r_{1}\!\mid\!\r_{2},\hat{\uu}_{2}) \right] \nonumber \\
&&\;\;\;\;\;\;\;\;\;\;\;\;\;\;\;\;\;\;\;\;\;\;\;\;\;\;\;\;\;\;\;\;\;\times\:
\exp\{i\mathbf{k}\cdot \r_{12}\} \,
j_{0}(\frac{1}{2}L\mathbf{k}\cdot \hat{\uu}_{1})\,
j_{0}(\frac{1}{2}L\mathbf{k}\cdot \hat{\uu}_{2})\:,
\end{eqnarray}
where,
\begin{eqnarray}
j_{0}(x)\:\equiv \: \frac{\sin\{x\}}{x}\:.
\end{eqnarray}
Consider the integral with respect to $\r_{12}$,
\begin{eqnarray}
I\:\equiv\:\int d\r_{12}\left[ g(\r_{12},\hat{\uu}_{1},\hat{\uu}_{2})-\chi_{f}(\r_{1}\!\mid\!\r_{2},\hat{\uu}_{2}) \right]
\exp\{i\mathbf{k}\cdot \r_{12}\}\:.
\end{eqnarray}
Replace the expression in the square brackets by $(g-1)+(1-\chi_{f})$. The integral
over $1-\chi_{f}$ is easily found to be equal to,
\begin{eqnarray}
\int d\r_{12} \left[ 1- \chi_{f}(\r_{1}\!\mid\!\r_{2},\hat{\uu}_{2}) \right]
\,\exp\{i\mathbf{k}\cdot \r_{12}\}
\,=\, \frac{\pi}{4}D^{2}L\,j_{0}(\frac{1}{2}L\mathbf{k}\cdot \hat{\uu}_{2})\:,
\end{eqnarray}
while the integral over $g-1$ is equal to,
\begin{eqnarray}
\int d\r_{12} \left[ g(\r_{12},\hat{\uu}_{1},\hat{\uu}_{2}) -1\right]
\,\exp\{i\mathbf{k}\cdot \r_{12}\}\,=\,-2DL^{2}
\mid\! \hat{\uu}_{1}\times \hat{\uu}_{2} \!\mid
j_{0}(\frac{1}{2}L\mathbf{k}\cdot \hat{\uu}_{1})
j_{0}(\frac{1}{2}L\mathbf{k}\cdot \hat{\uu}_{2})  \:.
\end{eqnarray}
These results are valid for $kD\ll 1$ (say $kD<0.2$), while, in addition, eq.(75) is
valid for orientations where
$\frac{D}{L}\ll \mid\!\hat{\uu}_{1}\times \hat{\uu}_{2}\!\mid$.
As will turn out, the $kD$-dependence is of no importance for long and thin
rods, since convergence of the wavevector integral is assured by the $kL$-dependent
functions, which tend to zero for wavevectors for which, indeed, $kD\ll 1$. Moreover,
the angular integration range, pertaining to orientations where
$\frac{D}{L}/\mid\!\hat{\uu}_{1}\times \hat{\uu}_{2}\!\mid$ is not small,
vanishes for long and thin rods. Substitution of the results (74,75) into
eq.(71) for $f_{1}$, and noting that after integration over orientations
the dependence on the direction $\hat{\mathbf{k}}$ of the wavevector is lost, so
that its direction may be chosen along the $z$-direction, yields (with $x=\frac{1}{2}kL$),
\begin{eqnarray}
f_{1}\,=\,\frac{2}{\pi^{4}}\int_{0}^{\infty} \!\!\!dx \oint d\hat{\uu}_{1}
\oint d\hat{\uu}_{2}\,
j_{0}^{2}(x\,z_{2})\,
\left[ \mid\!\hat{\uu}_{1}\times \hat{\uu}_{2}\!\mid
j_{0}^{2}(x\,z_{1})
-\frac{\pi}{8}\frac{D}{L} j_{0}(x\,z_{1})\right] .
\end{eqnarray}
with $z_{j}$, $j=1,2$, is the $z$-component of $\hat{\uu}_{j}$. The second
term between the square brackets is an $\mathcal{O}(D/L)$ contribution as
compared to the first term and may be neglected. Transforming the
orientational integrals to spherical coordinates, for which $z_{j}=\cos\{ \Theta_{j}\}$,
and using that (with $\Psi = \varphi_{1}-\varphi_{2}$),
\begin{eqnarray}
\mid\!\hat{\uu}_{1}\times \hat{\uu}_{2}\!\mid =\left[
1\,-\left( \cos\{\Theta_{1}\}\cos\{\Theta_{2}\}
+\sin\{\Theta_{1}\}\sin\{\Theta_{2}\}\cos\{\Psi\}\right)^{2}\right]^{1/2}\,,
\end{eqnarray}
finally yields eq.(46) for $f_{1}$.

Next consider the evaluation of the integrals in eq.(44) for $f_{2}$. That the
integrals are convergent follows from the Taylor expansion,
\begin{eqnarray}
\frac{1}{\mid\!\r_{12}-\mathbf{a}\!\mid}\,=\,
\frac{1}{r_{12}}+\mathbf{a}\cdot \nabla \frac{1}{r_{12}}
+\frac{1}{2}\mathbf{a}\mathbf{a}:\nabla\nabla \frac{1}{r_{12}}+\cdots \:.
\end{eqnarray}
Using this in eq.(44) and integration with respect to $\hat{\uu}_{1}$ shows
that the integrand varies like $\sim r_{12}^{-4}$ for large $r_{12}$, since
$\nabla^{2}r_{12}^{-1}=0$ for $r_{12}\neq 0$. Following the same procedure
as above one finds,
\begin{eqnarray}
f_{2}&=&\frac{1}{2\pi^{3}DL} \int d\mathbf{k} \int_{-1}^{1}dz_{1}
\int_{-1}^{1}dz_{2}\,\left[ (2\pi )^{3}\delta (\mathbf{k})
-\frac{\pi}{4}D^{2}L\,j_{0}(\frac{1}{2} Lk\,z_{2}) \right] \nonumber \\
&&\;\;\;\;\;\;\;\;\;\;\;\;\;\times \: j_{0}(\frac{1}{2}Lk\,z_{2})
\:\frac{j_{0}(\frac{1}{2}Lk\,z_{1})-1}{k^{2}}\:,
\end{eqnarray}
where $\delta$ is the delta distribution. The second term in the square brackets
is easily seen to be $\mathcal{O}(D/L)$, using the same integration tricks as
above for the evaluation of $f_{1}$. For the evaluation of the delta distribution
contribution, the integrand can be expanded in a power series expansion in $k$.
Using that $j_{0}(x)=1-x^{2}/6+\cdots$, results in eq.(47) for $f_{2}$.

\section*{Appendix C}

An analytical centrifuge measures the concentration of sedimenting
colloid along the centrifugal field. From a single run in an
analytical ultracentrifuge we obtain a time sequence of plots
usually taken every few minutes. A representative sequence of these
plots is shown in Fig. \ref{Time_Series}. Each plot in the series
indicates the {\it fd} concentration as a function of radial
position in the cell at that particular time. The concentrations of
the dilute virus solutions were determined with the  extinction
coefficients of 3.84 mg$^{-1}$cm$^2$ at 270 nm [7]. For samples
with higher concentration the solution is optically opaque at 270
nm and therefore we measure absorbance at progressively higher
wavelengths, which corresponds to a lower extinction coefficient of
{\it fd}. The sedimenting particles in Fig. \ref{Time_Series} move
from left to right. The water-air interface is indicated by a sharp
peak located at radial position of 5.95 cm that is due to
refraction by the meniscus. Note that this peak does not move as a
function of time indicating that the container does not leak. As
the rods start sedimenting towards the cell bottom, the region at
the top of the solution (to the right of the air-water interface
and to the left of the sedimentation front in Fig.
\ref{Time_Series}) is depleted of virus as indicated by absence of
absorbtion. Also the value of the concentration of rods in the
plateau region, always to the right of the depleted region, is
decreasing as the bulk of the sample moves towards the bottom of
the container. The reason for this is that the walls of the cell
are not parallel to each other, but instead follow the lines of
centrifugal field in order to minimize convective disturbances, an
effect refered to as ``radial dilution'' [15]. Between the flat
plateau region and the depleted region there is a sharp boundary.

At higher concentrations of {\it fd} we observed the appearance of
a sharp peak at the sedimenting boundary as is shown in Fig.
\ref{Time_Series_With_Peak}. The peak height increases with
increasing concentration while the magnitude of the peak is
independent of the wavelength and thus this peak cannot be due to
absorption of the {\it fd}, which is wavelength dependent.
 The probable cause of the peak is the refraction of incident light
due to the steep gradient in the virus concentration and hence the
refractive index at the sedimenting boundary. As the incident light
is refracted away from the detector, less light is collected by it
and this results in apparent increased absorption of the sample.
The peak at the water/air meniscus has the same origin.

Two factors that determine the shape of the sedimenting boundary
are the diffusion constant and the self-sharpening effect [17]. The
diffusion of the particles leads to gradual spreading of the
initially very sharp boundary. This diffusion of particles is
countered by the self-sharpening effect, which is due to the
concentration dependence of the sedimentation velocity. On one
hand, any molecule lagging behind the boundary is in a more dilute
environment and will therefore sediment at an enhanced velocity. On
the other hand, the particles in the plateau region are in a more
concentrated environment and their sedimentation will be retarded.
As a consequence the boundaries will self-sharpen. In a suspension
of elongated particles the self-sharpening effect will be much
stronger then in a suspension of globular particles because the
volume prefactor $\alpha$ in Eq. \ref{Dhont} is much larger for
elongated particles then for globular particles. The pronounced
self-sharpening effect leads to hyper-sharp boundaries, which
result in a steep gradient of refractive index which in turn causes
the artifacts shown in Fig. \ref{Time_Series_With_Peak}. In
globular colloids these effects are usually not observed.

In sedimentation analysis it is assumed that the rate of movement
of the sedimentation  boundary is approximately equivalent to the
sedimentation velocity of the particles in the plateau (bulk)
region. To compare results from different runs it is usual to
express the sedimentation velocity in units independent of
centrifugal force as follows:

\begin{equation}
\label{definition}
S=\frac{1}{\omega^2 r}\frac{dr}{dt}=\frac{1}{\omega^2}
\frac {d \ln r}{dt}
\end{equation}

\noindent The sedimentation velocity unit is called a Svedberg (S),
with 1 S = $[10^{-13}\mbox{sec}^-1]$. We define $r$ as the radial
position at the sedimentation boundary where the virus
concentration is equal to half the concentration of the plateau
region. This quantity is easily obtained from experimental data for
samples at low concentration. For samples at higher concentration,
where we observe a peak at the sedimenting boundary due to
refraction of light, we define $r$ as the radial position of the
highest point of the peak.  A typical plot of the logarithm of $r$
against $\omega^2 t$ used in the determination of the sedimentation
constant is shown in Fig. \ref{posvstime}. Surprisingly, we found
out that a linear function provided an inadequate fit to our data.
When the sedimentation data are collected between radial positions
of 6.1 cm and 6.8 cm a polynomial of second order fits the data
much better:

\begin{equation}
\label{fited}
\ln r=A+B\omega^2t+C\omega^4t^2
\end{equation}

\noindent  We introduce the experimentally observed sedimentation
constant $S^{r}$ by combining Eq. \ref{fited} and Eq.
\ref{definition}:

\begin{equation}
\label{slope}
S^{r}=\frac{\mbox{d} \ln r}{\mbox{d}\omega^2t}
=B+2C\omega^2t
\end{equation}

\noindent From this equation we see that the experimentally measured
sedimentation velocity is not a constant but depends on the
position of the measurement $r$ of equivalently time $\omega^2t$ at
which the sedimentation front is found at position $r$. The reason
for this unexpected behavior is not clear, but we assume it is an
instrumental artifact. There is no physical reason to believe that
the sedimentation velocity is a function of time or of radial
position in the cell. In table~1 we see that the coefficient C,
obtained when the quadratic polynomial in Eq. \ref{fited} is fitted
to data in Fig.~\ref{Svedberg}a is independent of concentration.
This is another indication that this artifact is due to the
instrument.

Theoretically, the constant $B$ in eq. (\ref{slope}) should be
equal to the concentration dependent Svedberg constant:

\begin{equation}
\label{SvedbergEq}
 S_{\phi}=S_0(1-\alpha \phi)
\end{equation}

\noindent and the constant $C$ should be zero. Instead, we have found that
the experimental Svedberg (eq.~\ref{slope}) is described by:

\begin{equation}
\label{slope1}
S^{r}= S_{\phi}+ \mbox{offset} = S_0 + \mbox{offset}- S_0 \alpha
\phi
\end{equation}

\noindent where ``offset'' depends on position in the centrifuge,
but is independent of colloid concentration. $S_0$ is the Svedberg
constant of the rods in the limit of zero concentration. However,
the value of slope

\begin{equation}
\label{derivative}
\frac{\mbox{d} S^{r}}{\mbox{d}\phi} = \alpha S_0
\end{equation}

\noindent is independent of radial position (or equivalently $\omega^2t$) where
we evaluate Eq. \ref{slope} as is shown in table \ref{coeff}. The
measurement artifact only introduces a position dependent offset in
the sedimentation velocity which affects the measured value of
$S_0$.  From a few measurements where we did not observe
measurement artifacts ($C = 0$) we obtained the value of $S_0=47$.
Since this is in good agreement with previous measurements we use
this value throughout our analysis [18].

It is important to note that the dependence of $S^r$ on position
$r$ shown in Fig. \ref{posvstime} is not due to the decreasing
concentration of rods in plateau, which in turn is due to radial
dilution. To show this we have made two measurements. In a first
measurement we evaluated the sedimentation velocity at the point
where the sedimenting boundary is close to the bottom of the
container. At this time, due to radial dilution, the plateau
concentration is about 70\% of the initial concentration. In the
second run our initial concentration was 70\% of the concentration
of rods in the first run. In this run we evaluated the
sedimentation velocity right at the beginning of the run. We find
that sedimentation velocities obtained in these two ways are vastly
different, which indicates that the systematic errors described are
not due to radial dilution.

\newpage

{\bf References}

$\star$ Author to whom the correspondence should be addressed

1. A.P. Philipse, Current Opinion in Coll. and Int. Sci. {\bf 2},
200 (1997).

2. G.K. Batchelor, J. Fluid Mech. {\bf 52} 245 (1972).

3. J. M. Peterson, J. Chem. Phys. {\bf 40} (1964) 2680.

4. J. Garc\'{i}a de la Torre, V.A. Bloomfield, Quarterly Reviews of
Biophysics {\bf 14}, 81 (1981).

5. D. M. E. Thies-Weesie, Sedimentation and liquid permeation of
inorganic colloids, PhD thesis, Universiteit Utrecht, 1995

6. S. Fraden, in {\it Observation, Prediction, and Simulation of
Phase Transitions in Complex Fluids}, edited by M. Baus, L. F. Rull
and J. P. Ryckaert (Kluwer, Dordrecht, 1995)

7. S. A. Berkowitz and L. A. Day, J. Mol. Biol. {\bf 102}, 531,
(1976).

8. L. Song, U. Kim, J. Wilcoxon and J. M. Schurr, Biopolymers, {\bf
31}, 547 (1991).

9. K. Zimmermann, H. Hagedorn, C. C. Heuck, M Hinrichsen and H.
Ludwig, J. Biol. Chem., {\bf 261}, 1653 (1986)

10. J. Sambrook, E. F. Fritsch and T. Maniatis, Molecular Cloning:
a laboratory manual, (Cold Spring Harbor Laboratory Press, New
York, 1989).

11. R. Buscall , J. W. Goodwin, R. H. Ottewill and Th. F. Tadros,
J. Colloid Interface Sci., {\bf 85}, 78 (1982).

12. D. M. E. Thies-Weesie, A. P. Philipse, G. Nagele, B. Mandl and
R. Klein, J. Colloid Interface Sci., {\bf 176}, 43 (1995).

13. J. K. G. Dhont, {\it An Introduction to Dynamics of Colloids},
(Elsevier Science, Amsterdam, 1996).

14. L. Onsager, Ann. N. Y. Acad. Sci. {\bf 51}, 627 (1949).

15. J. Tang and S. Fraden, Liquid Crystals {\bf 19}, 459 (1995).

16. M. Doi and S. F. Edwards, {\it The Theory of Polymer
Dynamics},(Clarendon Press, Oxford, 1986)

17. E. Schachman, {\it Ultracentrifugation in Biochemistry},
(Academic Press, New York, 1959).

18. J. Newman, H. L. Swinney, and L. A. Day, J. Mol. Biol. {\bf
116}, 593, 1977

\newpage

\begin{figure}
\centerline{\epsfig{file=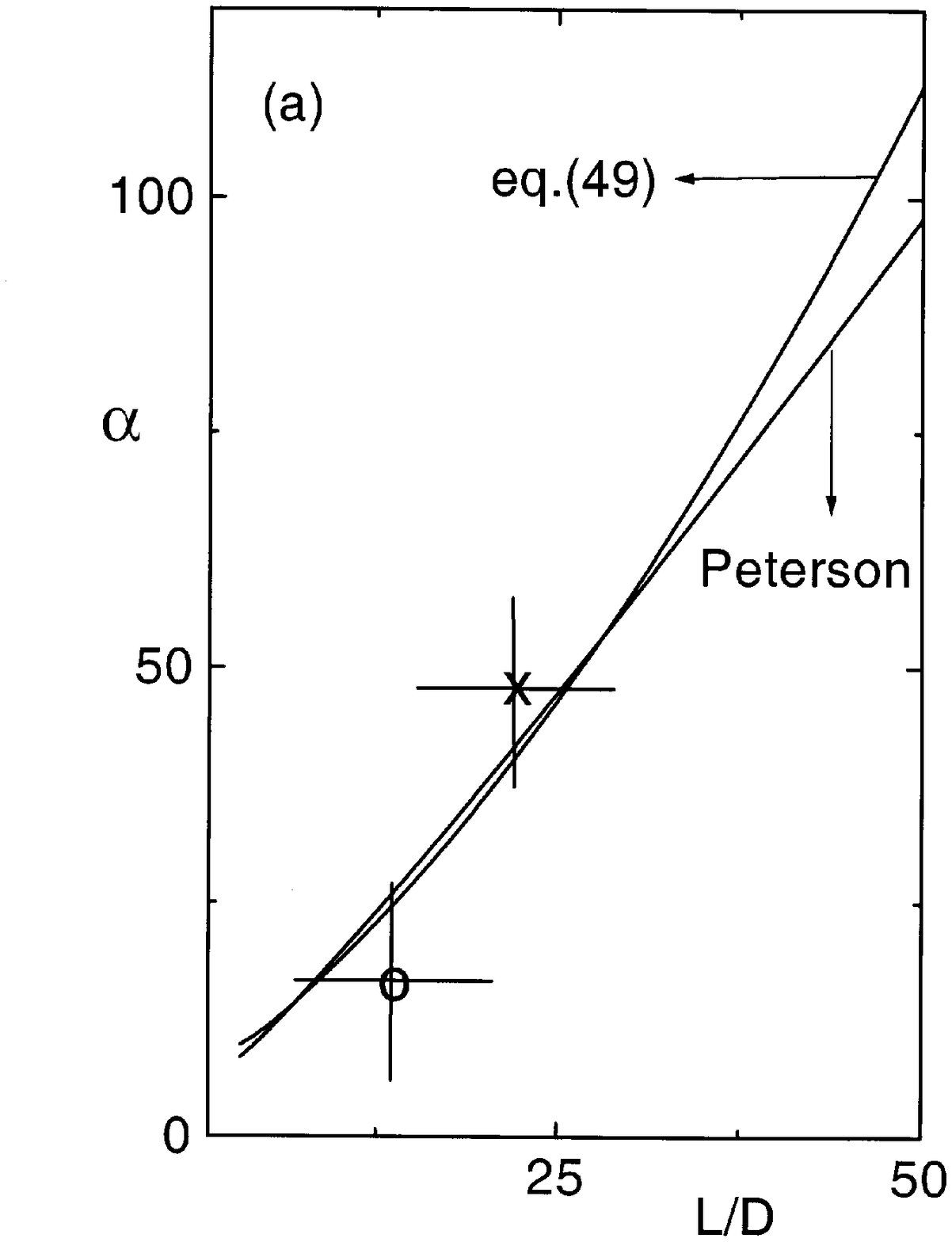,width=8.5cm}}
\centerline{\epsfig{file=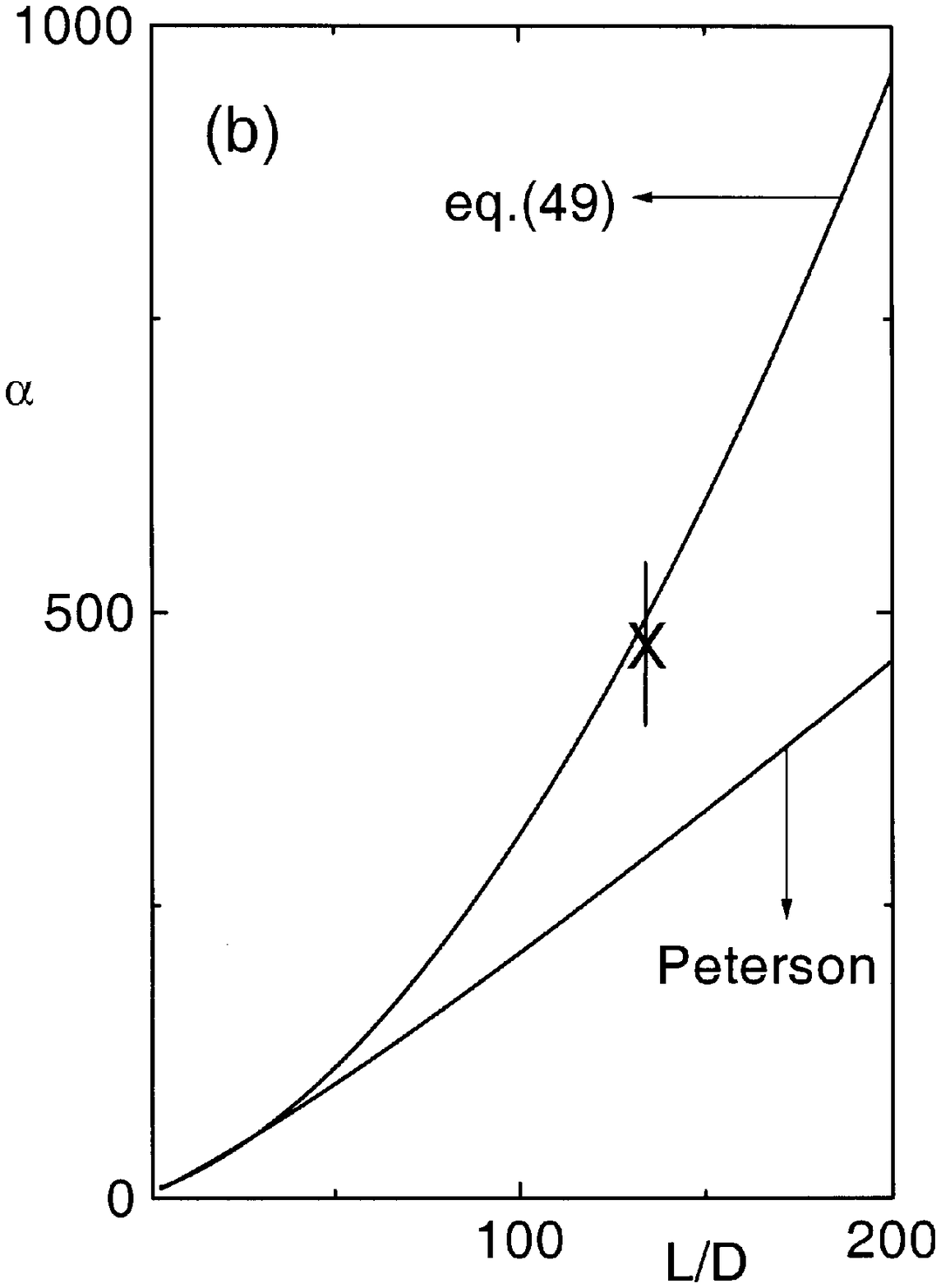,width=8.5cm}}
\caption{ a) The dependence of the coefficient $\alpha$ on L/D
ratio according to Eq. (49) and to Peterson [3]. The coefficient
$\alpha$ is the first order correction to sedimentation velocity
due to finite concentration of the colloidal rods. The two data
points are for silica rods. b) The coefficient $\alpha$ for
$L/D<200$. The data point is for $fd$-virus, as obtained in the
experimental section of the present paper}
\end{figure}

\newpage

\begin{figure}
\vspace{3cm}
\centerline{\epsfig{file=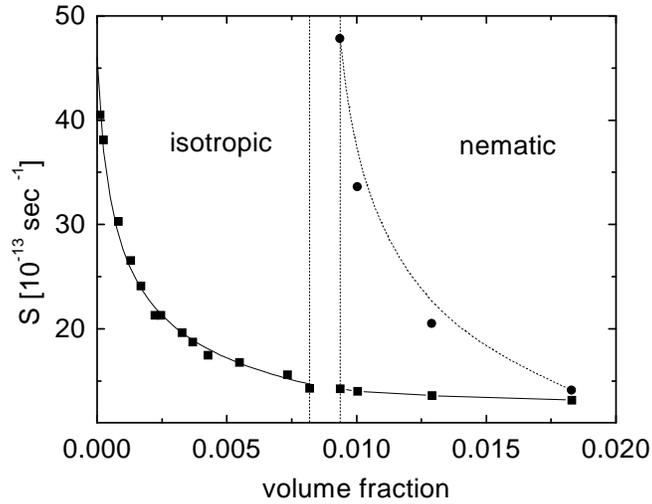,width=8.5cm}}
\caption{\label{10mMSvedberg}
Dependence of sedimentation velocity of rods measured in Svedberg's
(Eq.~\ref{slope}) on the average volume fraction of {\it fd} at an
ionic strength of 8 mM, pH 8.2. The equation of the fitted curve is
$S(\phi)=46.0(1+3600\phi)^{-1/3}$. A sample made with a volume
fraction between the two dashed vertical lines is unstable and will
spontaneously phase separate into an isotropic phase at volume
fraction of 0.0081 and a nematic phase at volume fraction 0.0093.
When rods are sedimented in the nematic phase of initial volume
fraction as indicated in the nematic region of the plot, we observe
two sedimenting boundaries with two different velocities and two
different concentrations. The sedimentation velocity of the faster
moving component is indicated with filled circles, while the
sedimentation velocity of slower moving boundary is indicated with
filled squares. The actual volume fraction of the slower moving
component is approximately constant at 0.0081 implying that that
component is in the isotropic phase. The concentration of the
faster component increases with the average concentration and is
always concentrated enough to be a nematic phase.}
\end{figure}

\newpage

\begin{figure}
\centerline{\epsfig{file=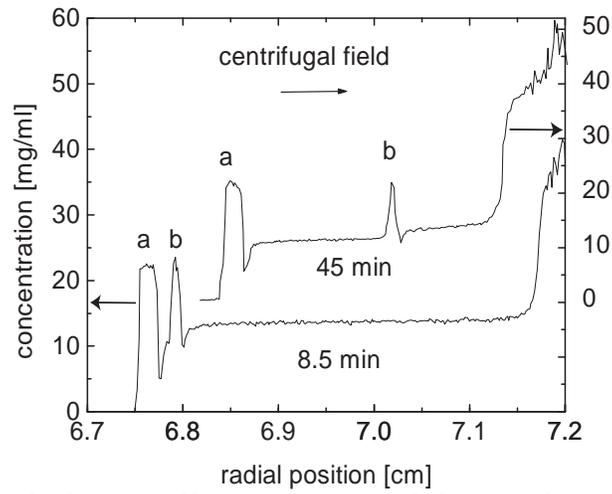,width=8.5cm}}
\caption{\label{Time_Series_Nematic} A concentration profile of sedimenting
{\it fd} virus in a nematic(cholesteric) phase taken at two
different times. Instead of a single moving boundary and a single
plateau we observe two moving boundaries and two plateaus.
Increased absorbance at the bottom of the container is due to the
accumulation of the virus particles. Peak ``b" marks the fast
sedimenting nematic boundary while peak ``a" marks the slow
sedimenting isotropic boundary. The two curves are offset for
clarity. The concentration of the initially uniform nematic sample
was 13 mg/ml. The concentration of the co-existing isotropic and
nematic phases at 8 mM ionic strength is 10.5 mg/ml and 12 mg/ml,
respectively.}
\end{figure}

\newpage

\begin{figure}
\centerline{\epsfig{file=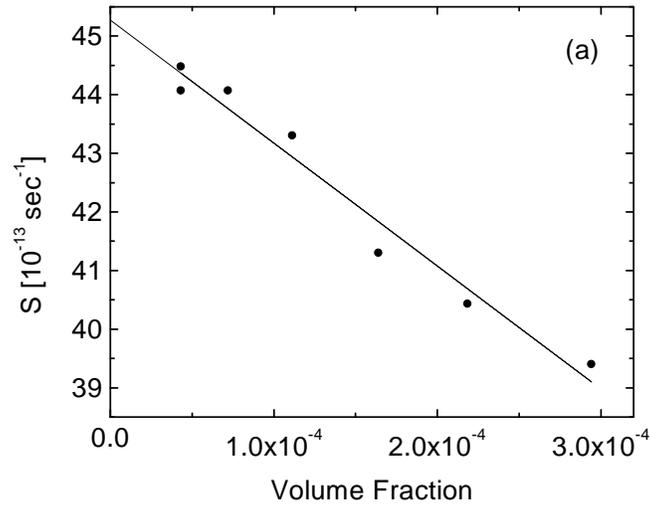,width=8.5cm}} \vspace{4cm}
\centerline{\epsfig{file=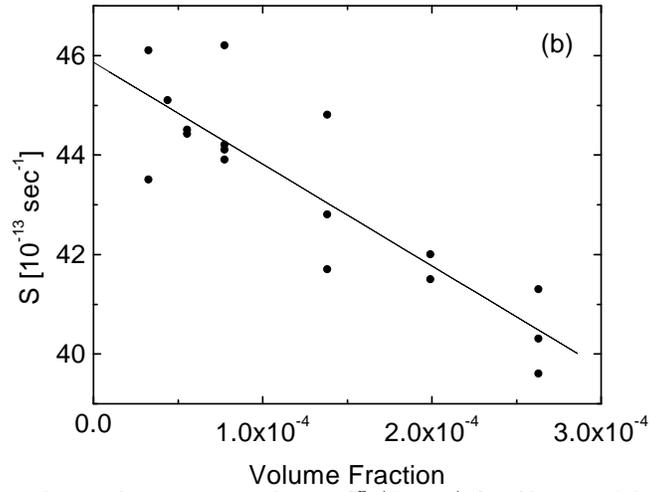,width=8.5cm}}
\caption{\label{Svedberg} a) Concentration dependent sedimentation
velocity $S^r$ (Eq.~\ref{slope1}) for {\it fd} at 50mM ionic
strength. The data is fitted to a linear function
$S^r=45.3-20980\phi$. The solid line is given by
Eq.~\ref{derivative}, which yields $\alpha S_0$. The overlap volume
fraction for {\it fd} with L=880 nm is $5.9\cdot 10^{-5}$ . b)
Concentration dependent sedimentation velocity for {\it fd} at
100mM ionic strength. The data is fitted to a linear function
$S^r=45.9-20450\phi$.}
\end{figure}

\newpage

\begin{figure}
\centerline{\epsfig{file=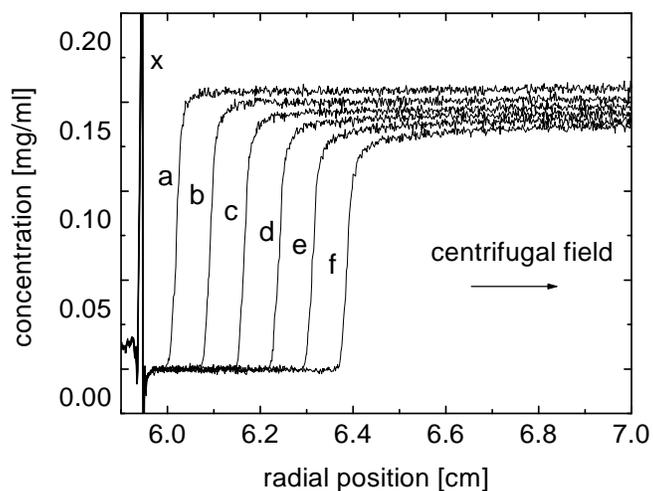,width=8.5cm}} \caption{\label{Time_Series}
Data obtained from an analytical centrifuge. A time series of the
{\it fd} concentration as a function of radial position in the
centrifuge taken at 6.5 min intervals with ``a'' the first scan and
``f'' the last. The steep step in concentration represents the
sedimentation front, which moves away from the centrifuge rotation
axis with time. In this particular case the centrifuge was spun at
25000 rpm and the centrifugal (sedimenting) field points from left
to right. The sharp peak ``x'' at the radial position of 5.95 cm is
due to refraction by the air-water meniscus. Radial dilution
accounts for the diminishing plateau concentration with increasing
time.}
\end{figure}

\newpage

\begin{figure}
\vspace{0.5in} \centerline{\epsfig{file=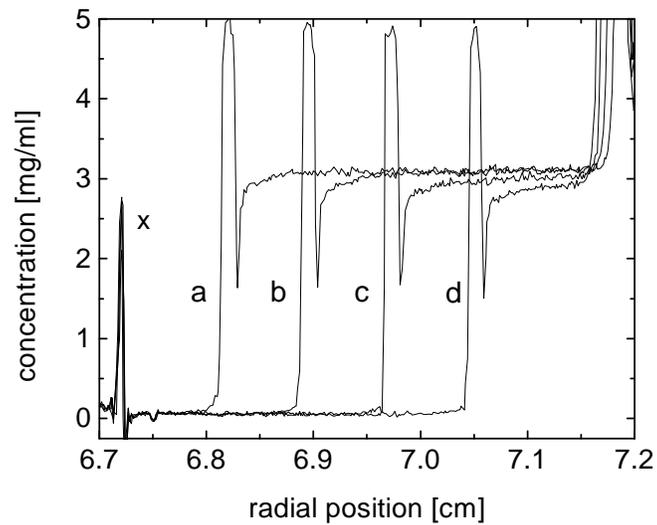,width=8.5cm}}
\caption{\label{Time_Series_With_Peak} A series of plots of {\it fd}
concentration as a function of radial position at time intervals of
approximately 12.7 min with ``a" the first scan and ``d" the last.
The difference between this series and those in Fig.
\ref{Time_Series} is that here the concentration of {\it fd} is
higher and the rotation speed was 20,000 rpm. The peak seen at the
sedimenting boundary is an artifact of the detection system and is
due to the refraction of light at a sharp step in the refractive
index at the sedimenting boundary. Radial dilution lowers the
plateau concentration with time. A similar peak occurs at the
air/water meniscus ``x", which is stationary.}
\end{figure}

\newpage

\begin{table}
\begin{tabular}{cccc} \hline
Volume fraction & S$^{6.1 \mbox{\scriptsize cm}}/10^{-13}$s$^{-1}$
& C & S$^{6.8 \mbox{\scriptsize cm}} /10^{-13}$s$^-1$\\
\hline
4.31$\cdot 10^{-5}$ & 44.1 & 1.35$\cdot10^{-23}$ & 49.5\\
4.31$\cdot 10^{-5}$ & 44.5 & 1.27$\cdot10^{-23}$ & 49.6\\
7.21$\cdot 10^{-5}$ & 44.1 & 1.21$\cdot10^{-23}$ & 48.9\\
1.11$\cdot 10^{-4}$ & 43.3 & 1.26$\cdot10^{-23}$ & 48.3\\
1.64$\cdot 10^{-4}$ & 41.3 & 1.17$\cdot10^{-23}$ & 45.9\\
2.18$\cdot 10^{-4}$ & 40.4 & 1.27$\cdot10^{-23}$ & 45.5\\
2.94$\cdot 10^{-4}$ & 39.4 & 1.35$\cdot10^{-23}$ & 44.8\\ \hline
\end{tabular}

\caption{\label{coeff} The values of constants obtained from Eq. \ref{slope}
being fitted to data from Fig. \ref{Svedberg}a. The second column
indicates the value of the sedimentation velocity $S^r$ evaluated
at the start of the sedimentation experiment at $\omega^2t=0$ or
equivalently r=6.1cm. The third column indicates the value of
parameter $C$ in Eq. \ref{slope}, which is independent of
concentration. If we evaluate Eq. \ref{slope} for the sedimentation
velocity at the end of the sedimentation experiment r=6.8 cm, we
obtain the values of the sedimentation velocity shown in the fourth
column. Note that the value of the slope $\alpha S_{0}$
(Eq.~\ref{slope1}) does not depend on the radial position. The
value of $\alpha S_{0}$ from the data evaluated at r=6.1cm is
20,500 and at r=6.8 cm is 21,000. We use the value $S_0 = 47$ to
obtain $\alpha$.}
\end{table}

\newpage

\begin{figure}
\centerline{\epsfig{file=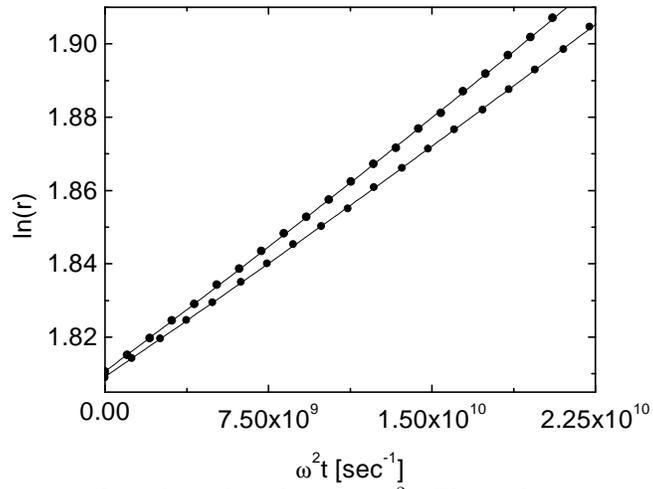,width=8.5cm}} \caption{\label{posvstime}
Position of the sedimentation boundary plotted against $\omega^2t$.
The circles represent measurements which were taken approximately
every 2.5 min. The lines represent the second order polynomial fit
to the data (Eq. 81). The plot with a larger slope corresponds to
the sedimentation of {\it fd} virus in 100mM ionic strength at a
volume fraction of 7.75$\cdot10^{-5}$ at 25,000 rpm. The other plot
is for a higher volume fraction of {\it fd} equal to
2.63$\cdot10^{-4}$. }
\end{figure}

\end{document}